\documentclass[12pt, final, onecolumn ]{IEEEtran}

\usepackage{graphicx}
\usepackage{color}

\usepackage{mathtools}
\usepackage{amsfonts}
\usepackage{amsmath}
\usepackage{setspace}
\usepackage{enumerate}
\usepackage{verbatim}
\usepackage{amssymb}
\usepackage{amsbsy}
\usepackage{theorem}
\usepackage{setspace}
\usepackage{url}
\usepackage[dvips]{epsfig}

 \newcommand{\sostfull}{contractive  after a small overshoot and short transient}
 \newcommand{\sostshort}{SOST}

\theoremstyle{plain}
\newtheorem{Theorem}{Theorem}

\newtheorem{Proposition}{Proposition}
\newtheorem{Corollary}{Corollary}

{\theorembodyfont{\rmfamily} \newtheorem{Example}{Example}}
\newtheorem{Remark}{Remark}

\newcommand {\R}{\mathbb R}

\newcommand{\be}{\begin{equation}}
\newcommand{\ee}{\end{equation}}
\newcommand{\Int}{\operatorname{{\mathrm int}}}

{\theorembodyfont{\rmfamily} }

\DeclareMathOperator{\pr}{Pr}

\newcommand{\LMD}{\lambda_0,\dots,\lambda_n}
\newcommand{\TLMD}{\tilde \lambda_0,\dots,\tilde \lambda_n}

\newcommand{\myc}[1]{{\bf{#1}}}  

\newcommand{\updt}[1]{#1}  

\newcommand{\model}{RFMEO}

\begin{document}
\title{Ribosome Flow Model with Extended Objects
\thanks{This research is partially supported by  research grants
 from  the Israeli Science Foundation, the Israeli Ministry of Science, Technology \& Space, \updt{the Edmond J. Safra Center for Bioinformatics at Tel Aviv University} and the US-Israel Binational Science Foundation.}
}
\author{Yoram Zarai,  Michael Margaliot and Tamir Tuller
\thanks{}}

\maketitle

\begin{abstract}

We study a deterministic mechanistic model for the flow of ribosomes along the  mRNA molecule, called
the \emph{ ribosome flow model with extended objects~({\model})}. This model
 encapsulates many realistic features of translation
 including non-homogeneous transition rates along the mRNA, the fact that every
ribosome covers several codons, and the fact that ribosomes cannot overtake one another.

The {\model} is a  mean-field approximation of an important model from
 statistical mechanics called the
\emph{totally asymmetric simple exclusion
 process  with extended objects~(TASEPEO)}. We demonstrate that the {\model}   describes biophysical aspects of translation better
than previous mean-field approximations, and that  its predictions
correlate well with those of~TASEPEO. However, unlike~TASEPEO, the {\model}
   is amenable to rigorous analysis using tools from systems and control theory.
We show
 that the ribosome density profile along the mRNA in the {\model}
 converges to  a unique steady-state density that depends on the length of the mRNA, the transition rates along it, and the
number of codons covered by every ribosome, but not on the initial density of ribosomes along the mRNA. In particular,
 the protein production rate also converges to a
unique steady-state.
Furthermore,  if the transition rates along the mRNA are periodic with a common period~$T$ then the ribosome density along the mRNA and the protein production rate converge  to a unique  periodic pattern with period~$T$, that is, the model entrains to periodic excitations in the transition rates.

Analysis and simulations of the {\model} demonstrate several   counterintuitive results.
For example, increasing   the ribosome footprint  may sometimes lead to an increase in the production rate. Also, for large values of the footprint  the steady-state density along the mRNA
may be quite complex (e.g. with quasi-periodic patterns) even for relatively simple (and non-periodic)
transition rates along the mRNA. This implies that inferring the transition rates from the ribosome density may be non-trivial.

We believe that the {\model}
could  be   useful   for modeling, understanding, and re-engineering translation as well as
 other important biological processes.

 \end{abstract}


\section{Introduction}
Gene expression is a multi-stage   process   for  converting
the  information inscribed in the DNA  to  proteins. During the transcription stage, the information   in the DNA of a specific gene is copied into messenger RNA~(mRNA). In the translation stage,
complex macro-molecules called ribosomes bind (at the initiation phase) to the mRNA and unidirectionally decode each codon (at the elongation phase) into the corresponding amino-acid that is delivered to the awaiting
ribosome by transfer RNA (tRNA). Finally, at the termination phase, the ribosome detaches from the mRNA, the amino-acid sequence is released, folded, and becomes a functional protein (in some cases, post-translation modifications may occur)~\cite{Alberts2002}. The output rate of ribosomes from the mRNA, which is also the rate in which proteins are generated, is called the protein translation rate, or production rate.

Translation occurs in all living organisms, and under almost all conditions. Thus, understanding the factors that affect translation has important implications to many scientific disciplines, including medicine, evolutionary biology, and synthetic biology. Deriving  and analyzing mechanistic  models of translation is  important for  developing a better understanding of this complex, dynamical, and tightly-regulated
 process. Such models can also aid in integrating and analyzing the rapidly increasing
 experimental findings related to translation~(see, e.g., \cite{Dana2011,TullerGB2011,Tuller2007,Chu2012,Shah2013,Deneke2013,Racle2013,Zur2016}).

Mechanistic  models of translation describe the dynamics of ribosome movement along the mRNA molecule, with   parameters that encode the various translation factors   affecting  the codon decoding times along the mRNA molecule. Several  such models have been suggested based on different
paradigms  ranging from Petri  nets~\cite{Brackley2012128}  to
probabilistic Boolean networks~\cite{Zhao2014}.
For more details, see the survey papers~\cite{Haar2012,Zur2016}.

The \emph{totally asymmetric simple exclusion process} (TASEP) \cite{Shaw2003,TASEP_tutorial_2011}
 is a fundamental model in non-equilibrium statistical mechanics that has been used to model numerous natural and artificial  processes~\cite{TASEP_book,Zur2016}, including ribosome flow during
  mRNA translation.
In TASEP, particles stochastically hop between consecutive  sites
along an ordered lattice of $N$ sites.
However, a particle cannot hop to an already occupied site.
 TASEP  encapsulates both the unidirectional flow of ribosomes along the mRNA molecule, and the \emph{interaction} between the  particles, as a particle in site~$i$ blocks the movement of a particle in site~$i-1$. This hard exclusion principle models particles that have ``volume'' and thus cannot overtake one other.
In the context of translation, the lattice represents the mRNA molecule, and the particles are the ribosomes.
The rate of hoping from site~$i$ to site~$(i+1)$  is denoted by~$\gamma_i$.
A particle can hop to [from] the first [last] site of the lattice at a rate~$\alpha$   [$\beta$].
The flow through the lattice converges to a steady-state value that depends on~$N$ and the vector of parameters:
\be\label{eq:tasep_rates}
\mu:=\begin{bmatrix} \alpha, \gamma_1,\dots,\gamma_{N-1},\beta \end{bmatrix}'.
\ee
The special case where all the internal hoping rates are assumed to be equal and normalized to one, i.e.~$\gamma_i:=1$, $i=1,\dots,N-1$, is referred to as the
  \emph{homogeneous TASEP (HTASEP)}.

In TASEP   a particle occupies a single site.
However, in translation
every  ribosome occupies not only the codon it is translating,
 but also codons after and before it. More precisely, the ribosome footprint is about~$10$ to~$11$
 codons, and its exit tunnel length is about~$31$ codons \cite{Alberts2002,Kaczanowska2007,Verschoor1998,Zhang2011,Ingolia2009}.
 In \emph{TASEP with  extended objects (TASEPEO)},  a particle occupies multiple sites along  the lattice~\cite{MacDonald1968,macdonald1969concerning,PhysRevE.76.051113,shaw2004mean,shaw2004local,lakatos2003,Shaw2003}. For TASEPEO with
 open-boundary conditions (i.e. where the two sides of the lattice
 are connected to two particle reservoirs, as assumed here)
  few rigorous analytical results  are known~\cite{PhysRevE.76.051113}. Mean-field approximations, domain-wall arguments, and extensive Monte Carlo simulations suggest
	that the \emph{homogeneous} TASEPEO converges to a steady-state, and that
	  the model has the same phase-diagram as~HTASEP, i.e. it contains three phases: low-density, high-density, and maximal current.
 The phase boundaries depend on  the extended object size~\cite{shaw2004local}. TASEPEO with two types of object sizes was studied in~\cite{ez2004effect}.
 It is important to mention that the extended objects concept  is relevant for other intracellular processes e.g.   transcription~\cite{edri2013,Rice1993,Kuhner2009}.

The \emph{ribosome flow model (RFM)}~\cite{reuveni} is a deterministic mathematical model for mRNA translation, obtained via a   mean-field approximation of~TASEP with open-boundary conditions. As such, it also inherits the property that the
particle size is equal to the  site size.
When the RFM is used to model translation based on real biological data, this issue
is handled by coarse-graining the mRNA molecule into sites composed
 of several consecutive codons. For every site the translation time of each codon in the site is
used to determine the translation time of the site in the~RFM (see e.g.~\cite{reuveni}).
It is not clear, however, how to systematically coarse-grain the~mRNA in a way that yields the best fidelity
between the model structure and parameters and the biological reality.

In this paper, we analyze for the first time a mean-field approximation   of~TASEPEO.
This is a deterministic model that we refer to as the
\emph{ribosome flow model with extended objects ({\model})}.
Using the theory of contractive dynamical systems, we rigorously prove that
the {\model} always converges to a steady-state.
In other words, the density profile of ribosomes along the mRNA molecule always
converges to a unique
steady-state, and thus so
does the protein production rate. This shows that the {\model}
 is robust in the sense that perturbations
(e.g. due to stochastic noise in the biochemical reactions)
 in the ribosome  density and production rate
die out with time. This also  means that we can reduce the problem of
studying the density profile and  protein production rate
to  studying the steady-state profile and production rate.
We also prove  that the {\model}
  entrains (or frequency-locks) to periodic excitations. We show using simulations that
	the {\model}, unlike the~RFM, correlates well with~TASEPEO.

The remainder of this paper is organized as follows. The next section briefly reviews the RFM. Section~\ref{sec:rfmeo} describes the {\model}. 
Section~\ref{sec:main} describes our main theoretical
results on the properties of the {\model}. Section~\ref{sec:compare}
studies the correlation between {\model}  and TASEPEO.
The final section summarizes and describes several directions for further research. \updt{To increase the readability of this paper, all the proofs are placed in Appendix~A.   Appendix~B describes 
 how the {\model} can be derived by a mean-field approximation of TASEPEO.}

\section{Ribosome Flow Model (RFM)}
The  RFM~\cite{reuveni} is a \emph{deterministic} model for mRNA translation that can be derived by a mean-field approximation of TASEP (see, e.g.,~\cite[section 4.9.7]{TASEP_book}, \cite[p. R345]{solvers_guide}  \updt{(see also Appendix~B in the special 
case where the extended object size is equal to  one site unit)}. In the RFM, mRNA molecules are coarse-grained into $n$ consecutive sites of codons. The state variable $x_i(t): \R_+ \to [0,1]$, $i=1,\dots,n$, describes the normalized ribosomal occupancy level at site $i$ at time $t$, where $x_i(t)=1$ [$x_i(t)=0$] indicates that site $i$ is completely full [empty] at time $t$. The model includes $n+1$ positive parameters that  describe the maximal possible
transition rate between the sites: the initiation rate into the chain $\lambda_0$, the elongation (or transition) rate from site $i$ to site $(i+1)$ $\lambda_i$, $i=1,\dots,n-1$, and the exit rate $\lambda_n$.

\begin{figure*}[t]
\centering
\includegraphics[width= 15cm,height=3cm]{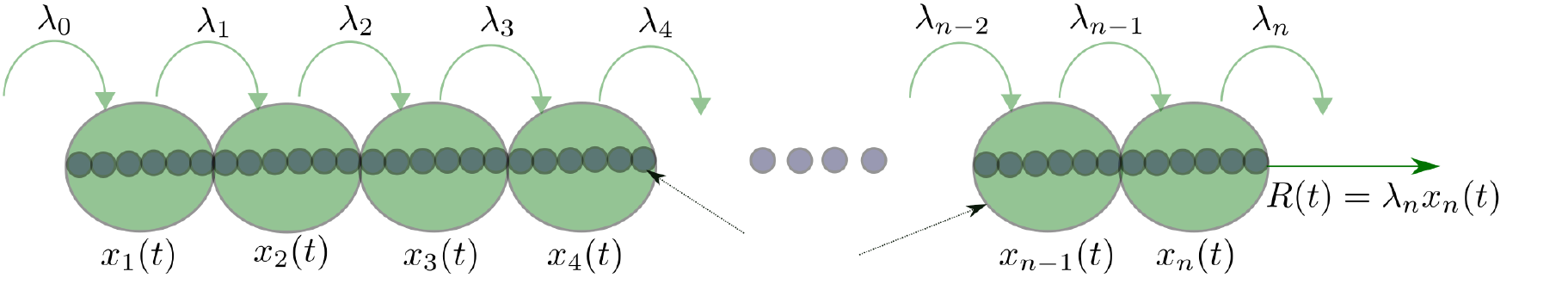}
\caption{The RFM as a chain of $n$ sites of codons. Each site is described by a state variable~$x_i(t)\in[0,1]$ expressing
 the normalized ribosome occupancy in site~$i$ at time~$t$. $\lambda_0$ is the initiation rate, and
 $\lambda_i$  is the elongation rate  from  site~$i$ to site~$(i+1)$. Production rate at time $t$ is $R(t):=\lambda_n x_n(t)$.}\label{fig:rfm}
\end{figure*}

The dynamics of the RFM with $n$ sites is given by $n$ nonlinear first-order ordinary differential equations:
\begin{align}\label{eq:rfm}
                    \dot{x}_1&=\lambda_0 (1-x_1) -\lambda_1 x_1(1-x_2), \nonumber \\
                    \dot{x}_2&=\lambda_{1} x_{1} (1-x_{2}) -\lambda_{2} x_{2} (1-x_3) , \nonumber \\
                    \dot{x}_3&=\lambda_{2} x_{ 2} (1-x_{3}) -\lambda_{3} x_{3} (1-x_4) , \nonumber \\
                             &\vdots \nonumber \\
                    \dot{x}_{n-1}&=\lambda_{n-2} x_{n-2} (1-x_{n-1}) -\lambda_{n-1} x_{n-1} (1-x_n), \nonumber \\
                    \dot{x}_n&=\lambda_{n-1}x_{n-1} (1-x_n) -\lambda_n x_n.
\end{align}

If we let $x_0(t):=1$ and $x_{n+1}(t):=0$,
then~\eqref{eq:rfm} can be written more succinctly as
\be\label{eq:rfm_all}
\dot{x}_i=h_{i-1}(x)-h_i(x),\quad i=1,\dots,n,
\ee
where $h_i(x):=\lambda_i x_i(1-x_{i+1})$.
This can be explained as follows. The flow of particles from site~$i$ to site~$(i + 1)$ at time~$t$ is $\lambda_i x_i(1-x_{i+1})$. This flow increases with the density at site $i$, and decreases as site $(i + 1)$ becomes fuller. This corresponds to a ``soft'' version of a simple exclusion principle: since the particles have volume, the input rate to site $i$ decreases as the number of particles in that site increases. Note that the maximal possible flow from site $i$ to site $(i + 1)$ is the transition rate $\lambda_i$. Thus Eq.~\eqref{eq:rfm_all}  simply states that the change in the density at site $i$ at time $t$ is the input rate to site $i$ (from site $i-1$) minus the output rate (to site $i + 1$) at time $t$.

The ribosome exit rate from site $n$ at time $t$ is equal to the protein production (or translation) rate at time $t$, and is denoted by $R(t):=\lambda_n x_n(t)$ (see Fig.~\ref{fig:rfm}). \updt{Note that $x_i$ is dimensionless, and that every rate~$\lambda_i$ has units of 1/time.}

Let~$x(t,a)$ denote  the solution of~\eqref{eq:rfm}
at time~$t \ge 0$ for the initial
condition~$x(0)=a$. Since the  state-variables correspond to normalized occupancy levels,
  we always assume that~$a$ belongs to the  closed $n$-dimensional
  unit cube:
\[
           C^n:=\{x \in \R^n: x_i \in [0,1] , i=1,\dots,n\}.
\]
Let~$\Int(C^n)$ denote the interior of~$C^n$, and let $\partial C^n$ denote the boundary of $C^n$.
It was shown in~\cite{RFM_stability} that if~$a\in C^n$ then~$x(t,a) \in C^n$ for all~$t\geq0$, that is,~$C^n$ is an invariant set of the dynamics. Ref.~\cite{RFM_stability} also showed that the RFM is a
\emph{tridiagonal cooperative dynamical system}~\cite{hlsmith},
and that this implies that~\eqref{eq:rfm}
admits a \emph{unique} steady-state 
  point~$e=e(\LMD) \in \Int(C^n)$, that is globally asymptotically stable, that is, $\lim_{t\to\infty} x(t,a)=e$, for all $a\in C^n$ (see also~\cite{RFM_entrain}). In particular, the production rate converges to the steady-state value~$R:=\lambda_n e_n$.

An important advantage of the RFM (e.g. as compared to TASEP) is that it is amenable to mathematical analysis using various tools from systems and control theory. Furthermore, most of the analysis results hold for the general, nonhomogeneous case (i.e. when  the transition
rates  all  differ from one another).
The RFM has been used to address many important biological problems including
the sensitivity of the production rate to small changes in the transition rate,
maximizing and minimizing the production rate in an optimal manner, analysis
of the effect of competition  for shared  resources in translation, and more~\cite{HRFM_steady_state,zarai_infi,RFM_stability, RFM_feedback, RFM_entrain, RFM_concave,RFM_sense, RFMR,rfm_control, RFM_r_max_density, Raveh2016,rfm_opt_down}.

To account for the fact that each ribosome covers several codons,
we analyze here the {\model}, which is a mean-field approximation of~TASEPEO (\updt{see Appendix~B for more details}). An
  integer~$\ell\ge1$ describes  the number of site units covered by each particle.
	The exclusion principle now implies that  the rate of flow from site~$i$ to site~$(i+1)$ is
		$\lambda_i x_i(1-x_{i+1}-\cdots-x_{i+\ell})$.
Indeed, since the particle  covers the next~$\ell$ sites, as  the density in any of  the~$\ell$
 consecutive sites increases the rate of movement slows down.
Note that~$\ell=1$ yields the RFM, so the RFM is a special case of the {\model}.

Nevertheless, the {\model} is a significant generalization of the~RFM and
its    dynamics   is quite different from that of the~RFM. For example, the {\model}, unlike the RFM, is \emph{not} a  cooperative system;
it does \emph{not}  satisfy the particle-hole symmetry of the RFM (and of TASEP)~\cite{rfm_opt_down,solvers_guide}, and
 unlike the RFM, the~{\model} with $\ell>1$ is \emph{not}  a tridiagonal dynamical
system.

\section{Ribosome Flow Model with Extended Objects ({\model})}\label{sec:rfmeo}
Being a large complex of molecules, each ribosome   typically covers
between $10$ to $11$ codons and the geometry (e.g. length of the exit tunnel) can be longer than 30 codons~\cite{Alberts2002}. A drawback of
 the~RFM and other standard mean field models
for translation is that, without additional processing such as coarse-graining,  each ribosome (``particle'') is assumed to cover a single site.

The {\model}
allows modeling the flow of ribosomes  where every ribosome covers
$1\le\ell\le n$   site units. We assume, without loss of generality, that the ribosome is translating the left-most site it is covering, and refer to this part of the ribosome as the \emph{reader}. A  similar assumption is used in~TASEPEO 
(see, for example,~\cite{PhysRevE.76.051113,shaw2004mean,shaw2004local,lakatos2003,Shaw2003,edge_tasep_2009}).
Thus, the statement ``the ribosome is at site~$i$''
 means that: the reader is located at site~$i$; the ribosome is translating site~$i$;
its corresponding transition rate is~$\lambda_i$; and sites~$i,\dots,i+\ell-1$, are covered by this ribosome.
As we will show below, the dynamical equations describing the~{\model} (and thus all the theoretical results in this paper) are the same for any chosen reader location (e.g. choosing the reader at location~$\ell/2$ results in exactly the same~{\model} equations).


Let $x_i(t)$ denote the (normalized) \emph{reader} occupancy level at site $i$ at time $t$,
and let $y_i(t)$ denote the (normalized) \emph{coverage} occupancy level at site $i$ at time $t$, that is,
\be\label{eq:y}
y_i(t):=\sum_{\mathclap{j=\max\{1,i-\ell+1\}}}^{i} x_{j}(t), \quad i=1,\dots,n.
\ee
Indeed, since every ribosome covers~$\ell$ sites, any ribosome that is located up to~$\ell$ sites left to site~$i$
contributes to the total ribosome coverage  at site~$i$.
\updt{The term ``normalized'' here means  that each $x_i(t)$ and each $y_i(t)$ takes values in the interval~$[0,1]$ for all $t\ge 0$. The value zero corresponds to   completely empty, and one means  completely full.}
We refer to~$1-y_i(t)$ as the  ``space'' or ``vacancy''
 level at site $i$ at time~$t$.

Note that~\eqref{eq:y}
implies that~$y(t)=Px(t)$, where~$P$ is the   lower triangular matrix  with all entries zero, except for the
entries on the  main diagonal and~$(\ell-1)$ diagonals below the main diagonal that are ones.
For example, for~$n=4$ and~$\ell=3$:
\be\label{eq:Pmat43}
					P=\begin{bmatrix}
					1 & 0 & 0 &0 \\
					1 & 1 & 0 &0 \\
					1 & 1 & 1 &0 \\
					0 & 1 & 1 &1
					\end{bmatrix} .
\ee

The dynamics of the {\model} with $n$ sites is given by $n$ nonlinear first-order ordinary differential equations:
\be\label{eq:rfmeo_ode}
\dot{x}_i =q_{i-1}(x)-q_i(x), \quad  i=1,\dots,n.
\ee
Here~$q_{i-1}$ is the flow  into site~$i$ and~$q_i$ is the flow
out of site~$i$.  The expression for this flow is given by
 \be\label{eq:rfmeo_flow}
q_i(x):=\lambda_i x_i (1-y_{i+\ell}),\quad i=0,\dots,n,
\ee

 with~$x_0(t)\equiv  1$, and $y_j(t)\equiv 0$  for all $j>n$.

\begin{figure}[t]
 \centering
 \includegraphics[width= 9cm,height=4cm]{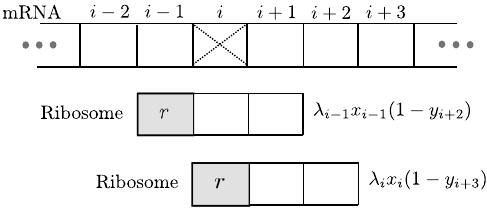}
\caption{Schematic explanation of the {\model} with $\ell=3$. Ribosomes that cover  three  sites scan the mRNA from  left-to-right. The label $r$ denotes the reader location. Shown are the two terms affecting the occupancy level at site $i$ in the {\model} dynamics (see~\eqref{eq:rfmeo_flow} and~\eqref{eq:rfmeo_ode}).}\label{fig:rfmeo_block}
\end{figure}

Eq.~\eqref{eq:rfmeo_flow} implies that the \emph{reader} flow from site $i$ to site $(i+1)$ is proportional to $\lambda_i$, to the occupancy levels of readers at site $i$, and to the ``space'' or ``vacancy''
 level at site $i+\ell$ (see Fig.~\ref{fig:rfmeo_block}). In particular,
\begin{itemize}
\item As the number of readers at site $i$ increases, the flow from site $i$ increases. This follows the same reasoning as in the RFM.
\item When a reader located at site $i$ moves to site $(i+1)$, the coverage occupancies at sites $i+1,i+2,\dots,i+\ell-1$ do not change. However,   the reader's tail end will now occupy a new site, which is site $(i+\ell)$.
\item The ``vacancy" level at site $(i+\ell)$ is $(1-y_{i+\ell})$, since $y_{i+\ell}$ denotes the total coverage at site $(i+\ell)$.
\end{itemize}

To explain~\eqref{eq:rfmeo_ode},
 consider for example the equation for the change in the density at site~$1$ given by
\begin{align*}
\dot{x}_1&=q_{0}(x)-q_1(x)\\
 &=\lambda_0   (1-y_{\ell})-\lambda_1 x_1 (1-y_{\ell+1}).
\end{align*}
The term~$\lambda_0   (1-y_{\ell})$ represents the entry rate into site~$1$.
Indeed, since  the entering ribosome will cover sites~$1,2,\dots,\ell$, this entry rate decreases
 with the coverage density~$y_\ell=x_1+\dots+x_\ell$.
(In the literature on  TASEPEO this is referred to  as the ``complete-entry'' flow~\cite{PhysRevE.76.051113}).
The term~$\lambda_1 x_1 (1-y_{\ell+1})$ is the flow from
 site~$1$ to site~$2$. This increases with the occupancy at site~$1$ and, similarly, decreases
with the coverage occupancy~$y_{\ell+1}$.

\begin{Remark}
As noted above, the assumption that the ``reading head''
 is located at the left hand-side  of the ribosome is arbitrary, but the {\model} equations do not depend on this assumption.
To demonstrate this, consider for example  the case~$\ell=3$ and the three possible locations for the reader:
(1) Left-most site. In this case $y_j=x_{j-2}+x_{j-1}+x_{j} $, so
\begin{align}\label{eq:ajfh}
q_i(x)=\lambda_i x_i(1-y_{i+3})=\lambda_i x_i(1-x_{i+1}-x_{i+2}-x_{i+3});
\end{align}
(2) Middle site. In this case $y_j=x_{j-1}+ x_j+x_{j+1}$, and $q_i(x)=\lambda_i x_i(1-y_{i+2})$,
and this yields~\eqref{eq:ajfh};
(3) Right-most site. In this case $y_j=x_j+x_{j+1}+x_{j+2}$, and $q_i(x)=\lambda_i x_i(1-y_{i+1})$,
again yielding~\eqref{eq:ajfh}.  Thus~\eqref{eq:rfmeo_flow} is invariant to the reader location.
\end{Remark}

Consider  an index~$j \ge n-\ell+2$. Then~$\ell+j-1  \ge n+1 $, so
\begin{align*}
\dot{x}_j&=q_{j-1}(x)-q_j(x)\\
 &=\lambda_{j-1}x_{j-1}   (1-y_{\ell+j-1})-\lambda_j x_j (1-y_{\ell+j})\\
 &=\lambda_{j-1}x_{j-1}    -\lambda_j x_j  .
\end{align*}
Thus, the equation describing the flow in these last
 sites is a linear equation. The same phenomena takes place in  TASEPEO, as
  a  ribosome ``reading''
 the last $\ell$ codons  must be the last particle on the lattice, with
no others to impede its progress. Therefore, it can move without hindrance
toward the exit end. The exit  rate in this context is referred to as the  ``incremental-exit''
rate~\cite{PhysRevE.76.051113}.

The output rate of ribosomes from the chain, which is the \emph{protein production (or translation) rate}, is denoted by $R(t):=\lambda_n x_n$.

Note that in the special case~$\ell=1$ we have $y_i=x_i$ for all $i=1,\dots,n$, and then~\eqref{eq:rfmeo_ode} reduces to
 the~RFM.
\begin{Example}
Consider a {\model} with dimension~$n=4$ and particle size~$\ell=2$. Then~\eqref{eq:rfmeo_ode} yields
\begin{align}\label{eqa:rfmeo42}
\dot{x}_1 & =  \lambda_0(1-y_2)-\lambda_1 x_1 (1-y_3), \nonumber \\
& = \lambda_0 (1-x_1-x_2) -\lambda_1 x_1(1-x_2-x_3), \nonumber\\
\dot{x}_2 & = \lambda_1 x_1(1-y_3)-\lambda_2 x_2 (1-y_4), \nonumber\\
& = \lambda_1 x_1 (1-x_2-x_3) -\lambda_2 x_2(1-x_3-x_4), \nonumber\\
\dot{x}_3 & = \lambda_2 x_2(1-y_4)-\lambda_3 x_3, \\
& = \lambda_2 x_2 (1-x_3-x_4) -\lambda_3 x_3, \nonumber\\
\dot{x}_4 &= \lambda_3 x_3 - \lambda_4 x_4.\nonumber
\end{align}
In the RFM the set~$C^n$ is an invariant set of the dynamics. This is no longer true for the~{\model}.
For example for the initial condition
\be\label{eq:inicon}
x(0)=\begin{bmatrix}  1&0.1&1&1 \end{bmatrix}',
\ee
Eq.~\eqref{eqa:rfmeo42} yields
\[
					\dot x_1(0)=0.1 (\lambda_1   -\lambda_0)  .
\]
This implies that~$x_1(0^+)>1$  for~$\lambda_1>\lambda_0$,
so~$x(0^+) \not \in C^n$.~$\square$
\end{Example}

\begin{Remark}
The Jacobian matrix of the dynamics~\eqref{eqa:rfmeo42}   is
\[
 J(x)= \begin{bmatrix}
  -\lambda_0-\lambda_1(1-x_2-x_3) & -\lambda_0 +\lambda_1 x_1 & \lambda_1 x_1 & 0 \\
  \lambda_1(1-x_2-x_3) &  -\lambda_1 x_1-\lambda_2(1-x_3-x_4) &  -\lambda_1 x_1 +\lambda_2 x_2 & \lambda_2 x_2 \\
  0 & \lambda_2(1-x_3-x_4) & -\lambda_2 x_2 -\lambda_3 & -\lambda_2 x_2 \\
  0 & 0 & \lambda_3 & -\lambda_4
 \end{bmatrix}.
\]
Note that there are off-diagonal entries here whose sign may change
with time (e.g. $-\lambda_0 +\lambda_1 x_1 $). This implies that the {\model},
unlike the RFM, is not a cooperative dynamical system.
\end{Remark}

It is useful to explicitly
 write the dynamics of the  {\model} in terms of the coverage state-variables
(i.e. the~$y$ state-vector). Recall that the proofs of all the results are placed in Appendix A.

\begin{Proposition}\label{prop:ydun}
The coverage state-variables in the {\model} satisfy:
\begin{align}\label{eq:rfmeo_ode_y}
\dot{y}_i=&\lambda_{0}(1-y_{\ell})-\lambda_i \left(\sum_{k=0}^{\lceil (i-\ell)/\ell\rceil} (y_{i-k\ell}-y_{i-k\ell-1})\right) (1-y_{i+\ell}), &&\quad 1\le i\le\ell,  \nonumber \\
\dot{y}_i=&\lambda_{i-\ell}  \left( \sum_{k=0}^{\lceil  ((i-\ell)/\ell)-1 \rceil} (y_{i-(k+1)\ell}-y_{i-(k+1)\ell-1})  \right)  (1-y_{i}) \nonumber \\
&-\lambda_i \left(\sum_{k=0}^{\lceil (i-\ell)/\ell \rceil} (y_{i-k\ell}-y_{i-k\ell-1})\right) (1-y_{i+\ell}), &&\quad  \ell<i\le n,
\end{align}
where~$\lceil z\rceil$ denotes  the smallest  integer that is larger than or equal to~$z$.
\end{Proposition}

\begin{Example}\label{exmp:n4l2}
Consider the {\model} with $n=4$ sites and particle size $\ell=2$. Then~\eqref{eq:rfmeo_ode_y}
yields
\begin{align*}
\dot{y}_1 & = \lambda_0(1-y_2)-\lambda_1 y_1 (1-y_3), \\
\dot{y}_2 & = \lambda_0 (1-y_2)-\lambda_2 (y_2-y_1) (1-y_4), \\
\dot{y}_3 & = \lambda_1 y_1 (1-y_3)-\lambda_3 (y_3-y_2+y_1), \\
\dot{y}_4 & = \lambda_2 (y_2-y_1) (1-y_4) - \lambda_4 (y_4-y_3+y_2-y_1).~\square
\end{align*}
\end{Example}


\section{Theoretical  Results}\label{sec:main}

\begin{figure}[t]
 \begin{center}
\includegraphics[width= 9cm,height=8cm]{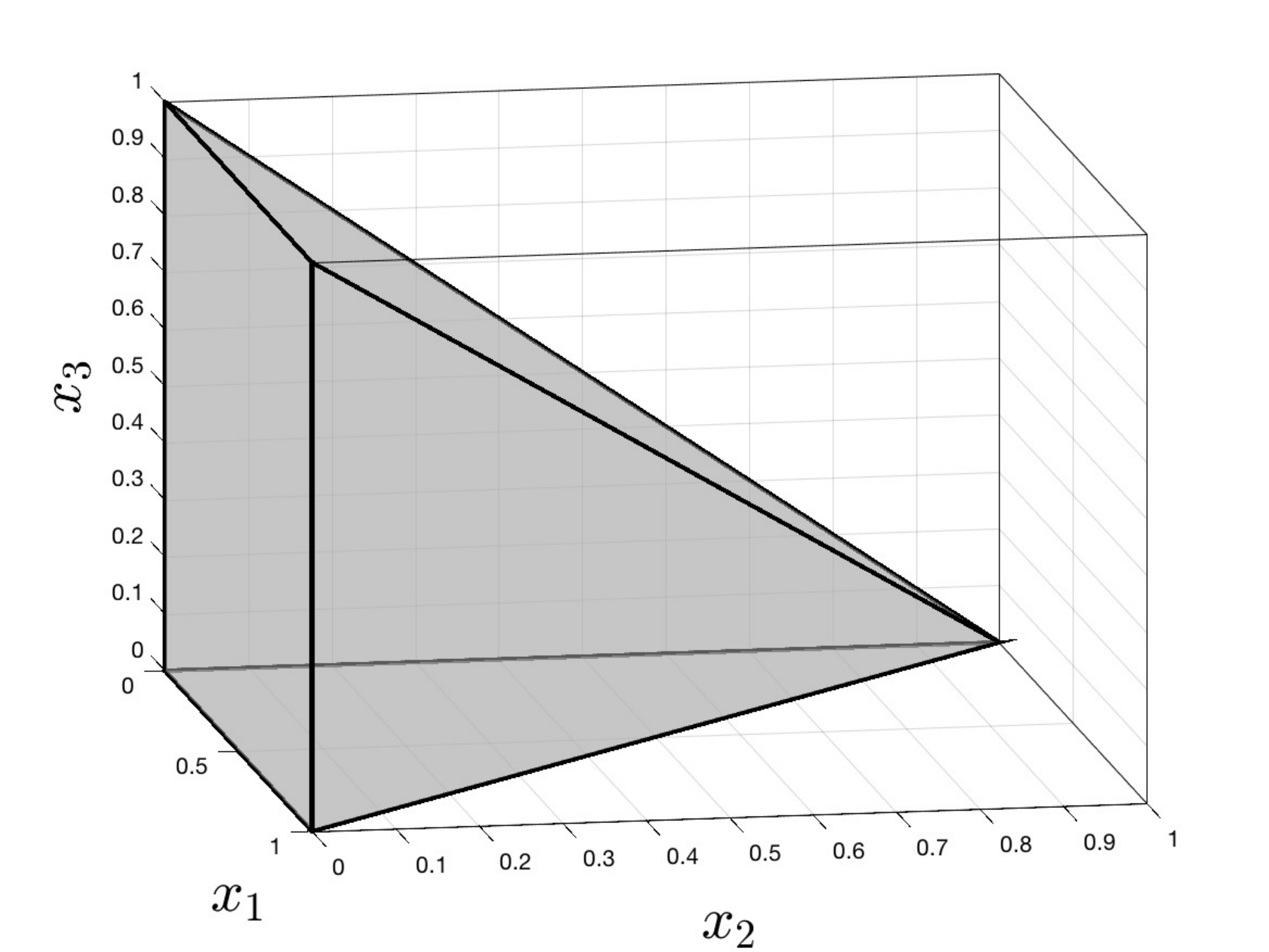}
\caption{{\model} with $n=3$ sites and particle size $\ell=2$. Gray volume is $H $, and the white cube is $C^3$. }\label{fig:n3l2_C_H}
 \end{center}
\end{figure}

We begin by defining    the relevant state-space for the {\model}.
If for some~$i$ we have~$y_{i+\ell}>1$ then~$q_i(x):=\lambda_i x_i (1-y_{i+\ell})<0$.
 This represents a backward flow that according to current knowledge does not take place  in
 ribosome movement. Thus, it is useful
to define the   state-space as the region where such a backward flow does not take place, i.e.
both the~$x_i$s and the~$y_i$s are between zero and one. This leads to defining
\[
  H  :=\{x\in \R^n  :x\in C^n  \text{ and }  P  x \in C^n\}.
\]
Note that~$H$ is a compact and convex set.
\begin{Example}
Consider the {\model} with $n=3$ sites and particle size $\ell=2$. The sets $H$ and~$C^3$ are depicted in Fig.~\ref{fig:n3l2_C_H}.~$\square$
\end{Example}

Note also that for
the {\model} with~$n=4$ and~$\ell=2$ the initial condition~$x(0)$ in~\eqref{eq:inicon} is
\emph{not} in~$H $ as~$y_2(0)=x_1(0)+x_2(0)>1$.

From here on we refer to any value~$x\in H$ as a \emph{feasible value}.
This  represents a state such
 that every reader density and every coverage
density is between zero and one.

\subsection{Invariance and persistence}

The next result shows that the boundary of~$H$, denoted~$\partial H$, is
``repelling'' \updt{towards the interior of $H$}. This means that if the dynamics is
initiated with a feasible value that includes
  a reader/coverage density equal to the extremal
value   zero [one]
then the dynamics will immediately change
 this to a value larger than zero [smaller than one].
\begin{Proposition}\label{prop:repel}
For any~$a \in \partial H$ the solution of the {\model} satisfies~$x(t,a)\in \Int(H)$ for all~$t >0$.
\end{Proposition}

Note that this result implies in particular that~$H$ is an invariant set of the dynamics.
In other words, if all the reader and coverage densities are initiated  with
 feasible values (i.e. values between zero and one) at time~$t=0$ then
they remain feasible for all time~$t\geq 0$.

The next result shows that the solutions of the {\model} are ``persistent'' in the sense that
they  enter and remain in a set that is  uniformly separated from the boundary of~$H$.
 Furthermore, this happens ``immediately''.
\begin{Proposition}\label{prop:persist}
 For any~$\tau>0$ there exists a compact and convex  set~$H_\tau$ that is \emph{strictly contained} in~$H$
such that for any~$a\in H$,
\[
				x(t,a) \in H_\tau, \text{ for all } t\geq \tau.
\]
\end{Proposition}

Note that this implies that for any~$\tau>0$
there exists~$\delta= \delta(\tau)  \in (0,1/2)$ such that
\[
  x_i(t,a),y_i(t,a) \in [\delta,1-\delta], \text{ for all } t\geq \tau,
\]
for all~$i$ and  all~$a\in H$.
In other words,  all
  the reader   and coverage
 densities    ``immediately'' become and remain uniformly
separated from the extreme values zero and one.
This is a technical property, but as we will see below it will be useful in the analysis of the  asymptotic
properties of the {\model}.

\subsection{Contraction}
Contraction theory is a powerful tool for analyzing nonlinear dynamical systems. In a contractive system, trajectories that emanate from different initial conditions approach
 each other at an exponential rate, that is,  the distance between any pair of trajectories, as a function of time, decreases at an exponential rate~\cite{LOHMILLER1998683,entrain2011,sontag_contraction_tutorial}.

 Consider the time-varying dynamical system
\be\label{eq:tvsys}
\dot{x}(t)=f(t,x(t)),
\ee
whose trajectories evolve on a compact and convex set~$\Omega \subset \R^n$.

\updt{
For~$t\geq t_0\geq 0$, and~$a\in \Omega$,
 let~$x(t,t_0,a)$ denote the
solution of~\eqref{eq:tvsys} at time~$t$ for the initial condition~$x(t_0)=a$.
Recall that
system~\eqref{eq:tvsys} is said to be contartive
  on~$\Omega$
w.r.t.  a norm~$|\cdot| :\R^n \to \R_+$ if there
    exists~$\gamma >0$
 such that
 \begin{align}\label{eq:ocont}
            |x(t_2,  t_1,a)-   x(t_2,t_1,b)|   \leq    \exp(-  (t_2-t_1)\gamma ) |a-b|
 \end{align}
for all $t_2\geq t_1\geq 0$  and all $a,b \in \Omega$. This means that any two trajectories approch each other at an
exponential rate~$\gamma$.
}

To apply contraction theory to the~{\model}, we require the following generalization of contraction with respect to a fixed norm that has been
 introduced in~\cite{3gen_cont_automatica}.
The time-varying
system~\eqref{eq:tvsys} is said to be
  \emph{\sostfull}~(\sostshort) on~$\Omega$
w.r.t.  a norm~$|\cdot| :\R^n \to \R_+$ if for each~$\varepsilon >0$
and each~$\tau>0$
  there exists~$\gamma=\gamma(\tau,\varepsilon)>0$
 such that
 \begin{align*}
            |x(t_2+\tau,& t_1,a)-   x(t_2+\tau,t_1,b)|   \leq   (1+\varepsilon) \exp(-  (t_2-t_1)\gamma ) |a-b| \,
 \end{align*}
for all $t_2\geq t_1\geq 0$  and all $a,b \in \Omega$.
\updt{Comparing this to~\eqref{eq:ocont}, we see that here
contraction ``kicks in'' after an arbitrarily small
time transient~$\tau$ and with an arbitrarily small overshoot~$(1+\varepsilon)$.   }

The next result applies these ideas to the~{\model}.
Let~$|\cdot|_1:\R^n\to\R_+$ denote the~$L_1$ norm, i.e. for $z\in\R^n$,~$|z|_1=|z_1|+\dots+|z_n|$.
\begin{Proposition}\label{prop:weak_cont}
The {\model} is {\sostshort} on~$H$
 w.r.t. the~$L_1$ norm, that is, for each~$\varepsilon >0$
and each~$\tau>0$
  there exists~$\gamma=\gamma(\tau,\varepsilon)>0$
 such that
 \begin{align}\label{eq:qcont}
            |x(t+\tau,  a)-   x(t+\tau, b)|_1   \leq
							(1+\varepsilon) \exp(-  \gamma  t ) |a-b|_1 \,
 \end{align}
for all $t \geq 0$  and all $a,b \in H$.
\end{Proposition}
Roughly speaking, this means the following.
 Fix two initial feasible densities in the {\model} and consider how the two
corresponding densities along the mRNA evolve in time.   Then these  densities become ``more similar'' to each other   at an exponential rate.  In particular, the initial  density is ``quickly forgotten''.

Subsections~\ref{subsec:stability} and~\ref{subsec:entrain} below describe important
asymptotic properties of  the {\model} that follow  from Prop.~\ref{prop:weak_cont}.


\subsection{Global asymptotic stability}\label{subsec:stability}
Write the   {\model}~\eqref{eq:rfmeo_ode}  as $\dot{x}=g(x)$, with~$g:H \to \R^n$.
 Since the compact and convex set~$H $ is an invariant set of this dynamical system, it contains a
 steady-state point $e=e(\LMD)$. In other words,~$g(e)=0_n$, where $0_n $ denotes a column vector of $n$ zeros,
and~$x(t,e)\equiv e$ for all~$t\geq0$.
Prop.~\ref{prop:repel} implies that~$e\in\Int(H )$. Using~\eqref{eq:qcont}
 with $b:=e$ yields the following result.

\begin{Corollary}\label{cor:asymp}
The {\model} admits a globally asymptotically stable steady-state point $e\in\Int(H )$, i.e.
\[
\lim_{t\to\infty} x(t,a)=e, \text { for all } a\in H .
\]
\end{Corollary}

This means that trajectories corresponding to different initial conditions all converge to the unique steady-state point, that depends on the transition rates~$\lambda_i$s,  particle size~$\ell$, and the length of the chain~$n$,
 but not on the initial condition.

\begin{figure}[t]
 \begin{center}
\includegraphics[width= 8cm,height=7cm]{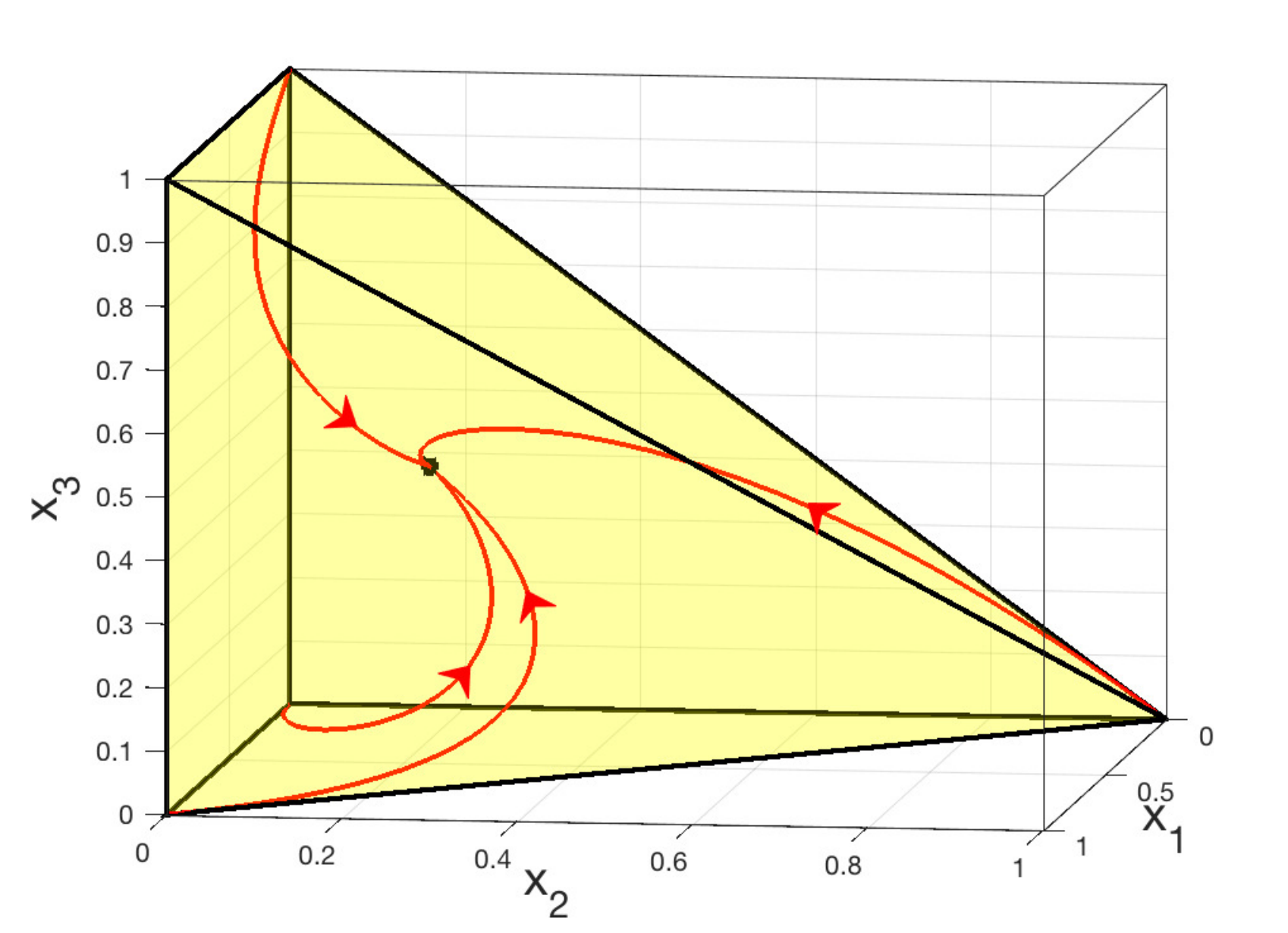}
\caption{Four trajectories of the {\model} with $n=3$, $\ell=2$, $\lambda_0=1.0$,
 $\lambda_1=1.2$, $\lambda_2=0.8$, and~$\lambda_3=0.4$. The unique steady-state point~$e\in \Int(H )$
is marked
 by a black dot.}\label{fig:n3l2_ss}
 \end{center}
\end{figure}

\begin{Example}
Fig.~\ref{fig:n3l2_ss} depicts the trajectories of an {\model} with dimension~$n=3$, particle size~$\ell=2$, and
rates~$\lambda_0=1.0$, $\lambda_1=1.2$, $\lambda_2=0.8$, and $\lambda_3=0.4$,
 for four different  initial conditions on the boundary of~$H$. It may be seen that each trajectory immediately enters and remains in the interior of~$H$,   and converges to a
 unique steady-state point~$e\in \Int(H )$.~$\square$
\end{Example}

The next example demonstrates the contraction property. Let $1_n$ denote the column vector of $n$ ones.

\begin{figure}[t]
  \begin{center}
  \includegraphics[scale=0.5]{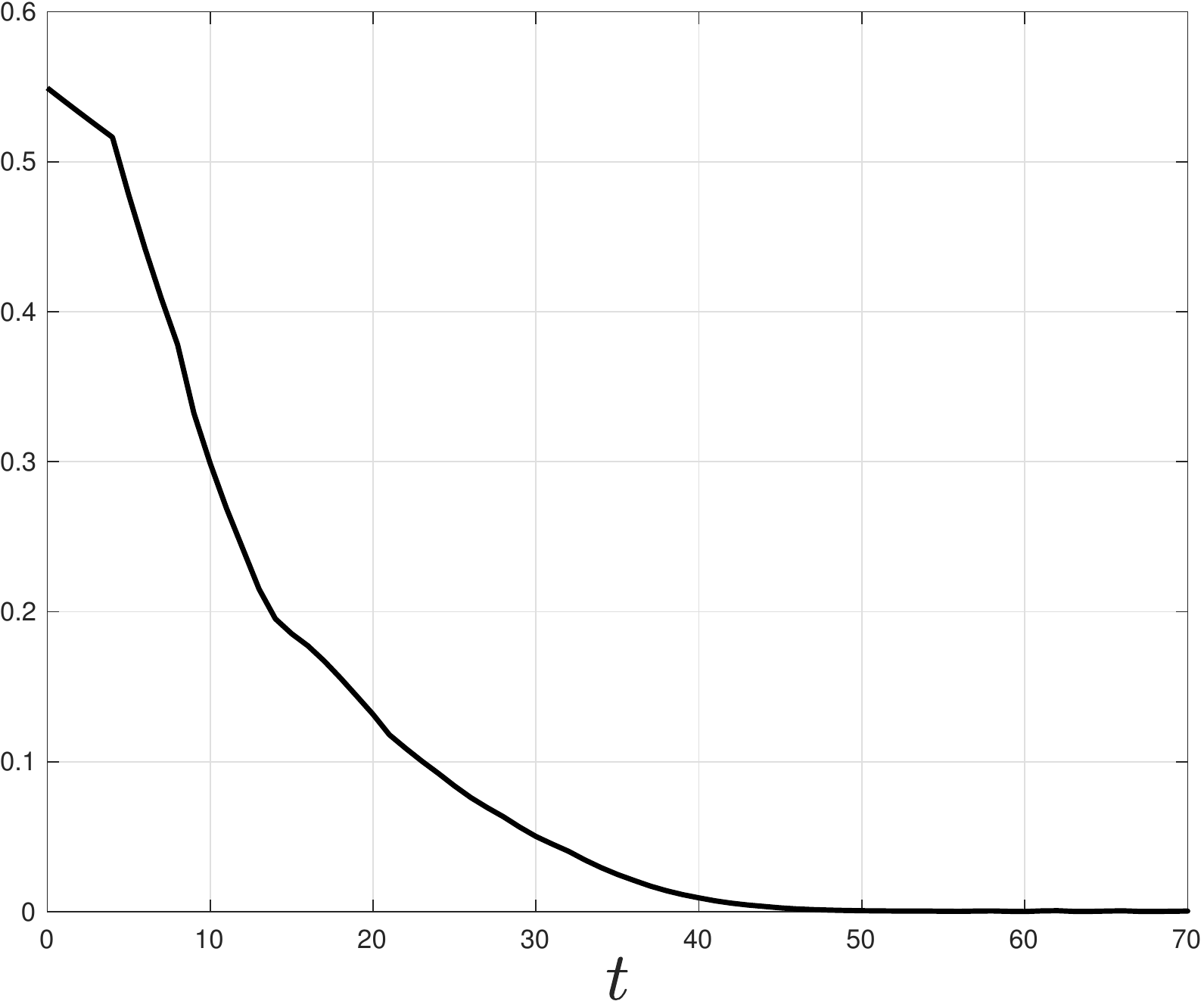}
  \caption{The distance~$|x(t,a)-e|_1$ as a function of~$t$
	for the {\model} in Example~\ref{exp:dist_e}.}\label{fig:rfmeo_n7_dist_e}
  \end{center}
\end{figure}

\begin{Example}\label{exp:dist_e}
Consider the {\model} with dimension~$n=7$, particle size~$\ell=3$,
and rates~$\lambda_i=1-\frac{i}{50}$, $i=0,\dots,7$. In this case the unique steady-state  point is (all numbers are to four digit accuracy):
\[
e =\begin{bmatrix} 0.3896  &  0.2697 &   0.2262 &   0.2042  &  0.1272  &  0.1302   &  0.1331\end{bmatrix}'.
\]
Fig.~\ref{fig:rfmeo_n7_dist_e} depicts $r(t):=|x(t,a)-e|_1$ as a function of time for $t\in[0,70]$ for
 the initial condition $a=(3/20)1_7$.
It may be seen that the $L_1$ distance between the trajectory and $e$ monotonically
decreases to zero. It may also be seen that the rate of convergence varies with time.
This makes sense because we can interpret the~{\model} as an~RFM with time-varying transition rates (see the proof of  Prop.~\ref{prop:weak_cont} in Appendix A), and thus
a time-varying contraction rate.~$\square$
\end{Example}

Corollary~\ref{cor:asymp} implies that
 the coverage occupancy~$y_i(t)$ at site~$i$ converges to the unique steady-state value:
\be\label{eq:z_eq}
z_i:=\sum_{\mathclap{j=\max\{1,i-\ell+1\}}}^{i} e_{j}, \quad i=1,\dots,n.
\ee

Define the \emph{mean reader occupancy} at time $t$ by
$
\rho(t):=\frac{1}{n}{\sum_{i=1}^n x_i(t)},
$
and the \emph{mean coverage occupancy} at time $t$ by
$
\rho^c(t):=\frac{1}{n} {\sum_{i=1}^n y_i(t)}.
$
 Note that this implies that $\lim_{n\to\infty} \rho^c(t)=\ell \rho(t)$.
Then the mean reader occupancy converges to the unique steady-state value
\be\label{eq:rfmeo_rho}
\rho:=\frac{1}{n} {\sum_{i=1}^n e_i},
\ee
and the mean coverage occupancy converges to the unique steady-state value
\be\label{rfmeo_rho_c}
\rho^c:=\frac{1}{n}\sum_{i=1}^n z_i.
\ee

\updt{The next example demonstrates the contraction property using
 a \emph{S. Cerevisiae} gene. Let~$0_n$ denote the column vector of $n$ zeros.}

\begin{figure}[t]
  \begin{center}
  \includegraphics[scale=0.5]{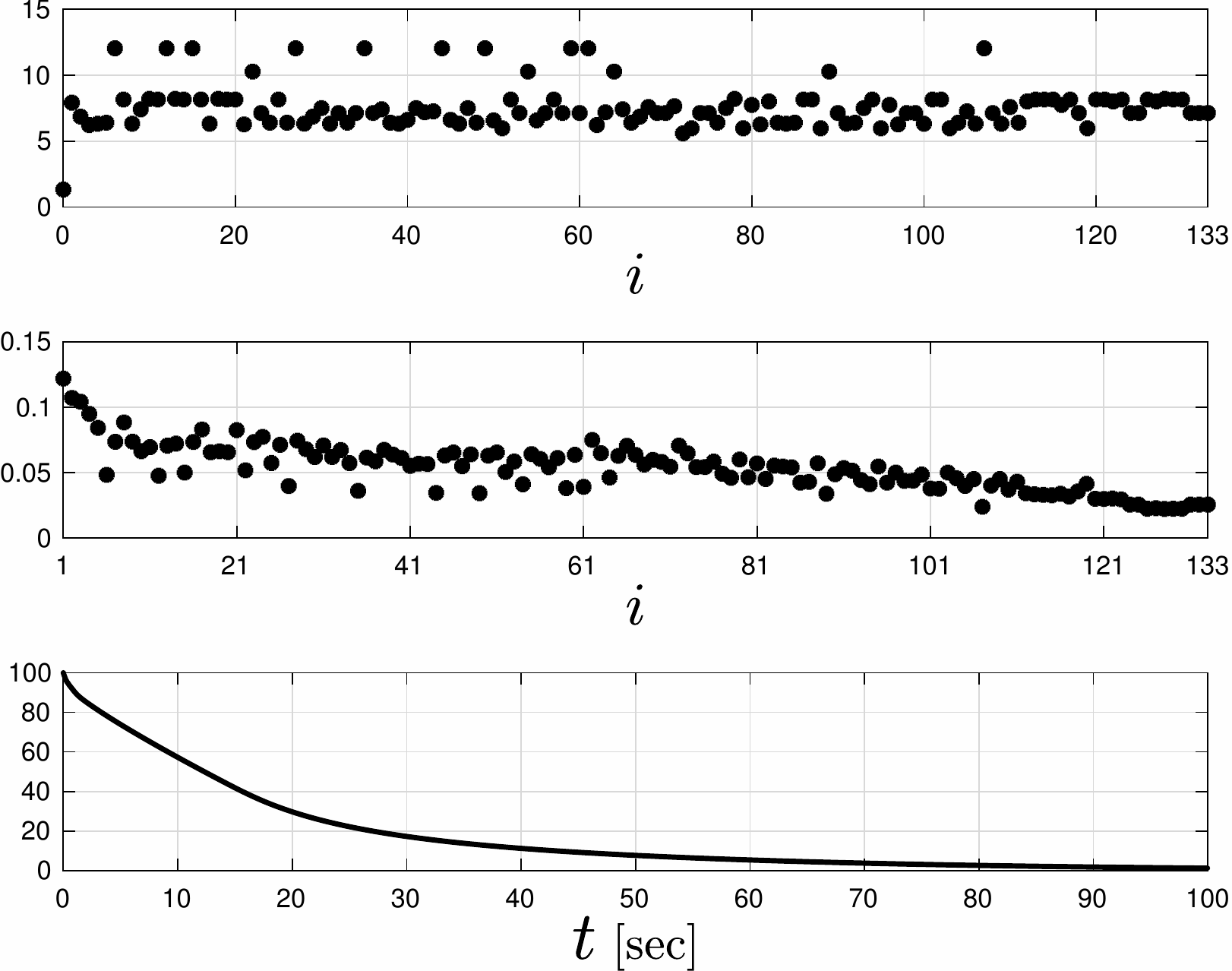}
  \caption{\updt{The biological model in Example~\ref{exp:scrv_dist}. Top: $\lambda_i$ as a function of $i$; Middle: steady-state reader density $e_i$ as a function of $i$; Bottom: relative $L_1$ distance   in percentage (see~\eqref{eq:rell1dd}) as a function of~$t$.}}\label{fig:scrv_e_dist}
  \end{center}
\end{figure}

\begin{Example}\label{exp:scrv_dist}
\updt{We consider the highly-expressed \emph{S. Cerevisiae} gene~YLR110C that encodes a  cell 
 wall mannoprotein, and  contains~$133$ codons (excluding the stop codon). We modeled it using a~{\model} with~$n=133$ and~$\ell=10$. The value $\lambda_0=1.33131$ [in units of mRNAs/sec] was estimated based
 on the ribosome density per mRNA levels, as this value is expected to be approximately proportional to
the initiation rate when initiation is rate limiting~\cite{reuveni,HRFM_steady_state}. 
The elongation rates $\lambda_1,\dots,\lambda_n$, were estimated using
   ribo-seq data for the
 codon decoding rates~\cite{Dana2014B}, 
 normalized so that the median elongation  rate of all {\em S. cerevisiae} mRNAs
becomes~$6.4$ codons per second \cite{Karpinets2006}. The  rates are depicted in the top panel of Fig.~\ref{fig:scrv_e_dist} as a function of~$i$.  
To study the rate of contraction, we calculated~$e$ in the~{\model} (shown in the middle panel of Fig.~\ref{fig:scrv_e_dist}),
and simulated the dynamical system to obtain
$r(t)=|x(t,a)-e|_1$ with~$a=0_{133}$, as a function of~$t$ [in  seconds]. Note that the
  initial condition~$a=0_{133}$   represents an~mRNA with no ribosomes.
The bottom panel of Fig.~\ref{fig:scrv_e_dist} depicts the relative~$L_1$ distance in percentage, that is,
\be\label{eq:rell1dd}
100 \frac{r(t)}{r(0)},
\ee
  as a function of~$t$ [in  seconds].
  In this case,~$\rho=0.0534$. It may be seen that the relative distance is less than~$20\%$ already after
	about~$30$ seconds. We note that typical \emph{S. Cerevisiae} mRNAs half-lives is in the order of tens of
	minutes (see, for example,~\cite{Shalem2008,Wang2002,Edri2014}). This suggests that typically the ribosome density
	on \emph{S. Cerevisiae} mRNAs is ``very close'' to its steady-state value.
}
\end{Example}

\subsection{Analysis of the steady-state}
It is important to understand how the steady-state density~$e$ depends on the parameters of the~{\model}.
To study this we begin by deriving some equations for~$e$.
At steady-state (i.e. for $x=e$), the left-hand side of all the equations in~\eqref{eq:rfmeo_ode} is zero (i.e. $\dot{x}_i=0$, $i=1,\dots,n$), so
\[
R = q_i(e)=\lambda_i e_i (1-z_{i+\ell}), \quad i=0,\dots,n,
\]
where $z_j:=0$, for all $j>n$. This yields (see~\eqref{eq:z_eq})
\begin{align}\label{eq:ss_e}
\lambda_0(1-e_1-\dots-e_{\ell})&=\lambda_1 e_1 (1-e_2-\dots-e_{\ell+1}) \nonumber \\
& = \lambda_2 e_2 (1-e_3-\dots-e_{\ell+2}) \nonumber \\
\vdots \nonumber \\
&=\lambda_{n-\ell-1} e_{n-\ell-1} (1-e_{n-\ell}-\dots-e_{n-1}) \nonumber \\
&=\lambda_{n-\ell} e_{n-\ell} (1-e_{n-\ell+1}-\dots-e_{n}) \nonumber \\
&=\lambda_{n-\ell+1} e_{n-\ell+1} \nonumber \\
\vdots \nonumber \\
&=\lambda_n e_n \nonumber \\
&=R.
\end{align}

 \updt{We can express these in terms of the (generally unknown value) $R$ as:}
\be\label{eq:ss_e_i}
e_i=\frac{R\lambda_i^{-1}}{1-e_{i+1}-e_{i+2}-\dots-e_{i+\ell}}=\frac{e_n \lambda_n\lambda_i^{-1}}{1-e_{i+1}-e_{i+2}-\dots-e_{i+\ell}},
\ee
and this yields
\begin{align*}
e_{n-j} &= R\lambda_{n-j}^{-1},\quad \quad \quad  j=0,\dots,\ell-1,\\
e_{n-\ell} &= \frac{R\lambda_{n-\ell}^{-1}}{1-R\sum_{k=0}^{\ell-1} \lambda_{n-k}^{-1}}, \\
e_{n-\ell-1} &= \cfrac{R\lambda_{n-\ell-1}^{-1}}{1-\cfrac{R\lambda_{n-\ell}^{-1}}{1-R\sum_{k=0}^{\ell-1} \lambda_{n-k}^{-1}}-R\sum_{k=1}^{\ell-1} \lambda_{n-k}^{-1}},\\
\vdots
\end{align*}
and
\[
0=1-\frac{R\lambda_0^{-1}}{1-e_1-\dots-e_\ell}.
\]

\begin{Example}
Consider the {\model} with dimension~$n=6$ and particle size~$\ell=3$. Then, the steady-state production rate $R$ satisfies
\[
0=1-\frac{R\lambda_0^{-1}}{1-e_1-e_2-e_3},
\]
where
\begin{align*}
e_1&=\cfrac{R\lambda_1^{-1}}{1-\cfrac{R\lambda_2^{-1}}{ 1-\cfrac{R\lambda_3^{-1}}{ 1-R(\lambda_4^{-1}+\lambda_5^{-1}+\lambda_6^{-1})  } -R(\lambda_4^{-1}+\lambda_5^{-1})   } -\cfrac{R\lambda_3^{-1}}{1-R(\lambda_4^{-1}+\lambda_5^{-1}+\lambda_6^{-1})} -R\lambda_4^{-1}   }, \\
e_2&=\cfrac{R\lambda_2^{-1}}{ 1-\cfrac{R\lambda_3^{-1}}{ 1-R(\lambda_4^{-1}+\lambda_5^{-1}+\lambda_6^{-1})  } -R(\lambda_4^{-1}+\lambda_5^{-1})}, \\
e_3&=\cfrac{R\lambda_3^{-1}}{ 1-R(\lambda_4^{-1}+\lambda_5^{-1}+\lambda_6^{-1})}.~\square
\end{align*}
\end{Example}

\updt{
It is clear that solving~\eqref{eq:ss_e} is in general non-trivial. Nevertheless, it can be
solved in closed-form in some very special cases. The next example demonstrates this.
\begin{Example}\label{exa:trac}
Consider a {\model} with~$n$ sites and with ribosome size~$\ell=n$. Then~\eqref{eq:ss_e} becomes
\begin{align}\label{eq:steady_l=n}
\lambda_0(1-e_1-\dots-e_{n})&=\lambda_1 e_1   \nonumber \\
& = \lambda_2 e_2   \nonumber \\
\vdots \nonumber \\
&=\lambda_n e_n \nonumber \\
&=R,
\end{align}
and this yields
\be\label{eq:poeq}
			e_i=\frac{1}{z} \prod_{\substack{j=0\\ j\not = i }}^{ n} \lambda_j   ,\quad i=1,\dots,n,
\ee
and
\be\label{eq:lkff}
R=\frac{1}{z}  \prod_{j=0  }^{ n} \lambda_j ,
\ee
where~$z:=\sum_{0\leq i_1<i_2<\dots<i_n\leq n} \lambda_{i_1}\lambda_{i_2}\dots \lambda_{i_n}$.
To understand this, assume in addition that~$\lambda_0=\dots=\lambda_n=\lambda_c$, i.e. all the rates are equal with~$\lambda_c$  denoting the common value. Then~\eqref{eq:poeq} and~\eqref{eq:lkff} yield~$e_i=1/(n+1)$ for all~$i$, and
\be\label{eq:rlm}
R=\lambda_c/(n+1).
\ee
This means that when the ribosome size is equal to the chain size and all the rates are equal
then the steady-state density at each site is identical. This makes sense, as every ribosome covers all the sites in the chain.

Another tractable case is when~$\ell=n-1$ and~$\lambda_0=\dots=\lambda_n=\lambda_c$.
Then~\eqref{eq:ss_e} yields
\begin{align}\label{eq:steady_l=nm1}
\lambda_c(1-e_1-\dots-e_{n-1})&=\lambda_c e_1  (1-e_2-\dots-e_{n}) \nonumber \\
& = \lambda_c e_2   \nonumber \\
\vdots \nonumber \\
&=\lambda_c e_n \nonumber \\
&=R,
\end{align}
and this admits the solution
\[
e_1=\frac{2}{\sqrt{4n-3}+1},\quad e_i=\frac{2}{\sqrt{4n-3}+2n-1} \text{ for all } i>1,
\]
and
\be\label{eq:rlm1}
R=  \frac{2\lambda_c}{\sqrt{4n-3}+2n-1}.
\ee
Note that here~$e_1>e_2=e_3=\dots=e_n$. This makes sense, because if there is a ribosome with 
reader at site~$\geq 2$ then the 
tail of this~$(n-1)$-sites long ribosome is either at site~$n$ or already out of the chain, and so there is no hindrance for its movement.~$\square$
\end{Example}
}

Eq.~\eqref{eq:ss_e} can also be used to prove theoretical results.
The next result shows that increasing any of the~$\lambda_i$s increases~$R$.
In other words,   increasing any of the transition rates along the mRNA molecule increases the steady-state protein production rate.

\begin{Proposition}\label{prop:R_mono}
Consider the {\model} with dimension $n$ and particle size $\ell$. Then $\frac{\partial}{\partial \lambda_i} R >0$, for $i=0,\dots,n$.
\end{Proposition}

In the special case where \emph{all} the    rates are equal, i.e.
\be\label{eq:thrfmeo}
\lambda_0=\cdots=\lambda_n:=\lambda_q,
\ee
where $\lambda_q$ denotes the  common value, we refer to the {\model}
as the  \emph{totally homogeneous {\model} (TH{\model})}.
  In this case, it is possible to say more about the steady-state occupancies.

\begin{Proposition}\label{prop:thrfmeo_e}
Consider the TH{\model} with dimension~$n$ and particle size~$\ell$. Then
\begin{align}\label{eq:thrfmeo_e}
&e_1>e_2>\cdots>e_{n-\ell+1}, \nonumber \\
& e_{n-\ell+1}=e_{n-\ell+2}=\cdots=e_n, \nonumber \\
& z_\ell>z_{\ell+1}>\cdots> z_n.
\end{align}
\end{Proposition}
This means that the steady-state reader occupancies monotonically decrease between sites $1$ and $(n-\ell+1)$ and are equal at the last $\ell$ sites. This may partially explain the decrease in ribosome density observed along the coding sequences from the~5' end to  the~3' end (see, for example, \cite{Dana2012,Ingolia2009}).

\begin{figure}[t]
 \begin{center}
\includegraphics[width= 8cm,height=7cm]{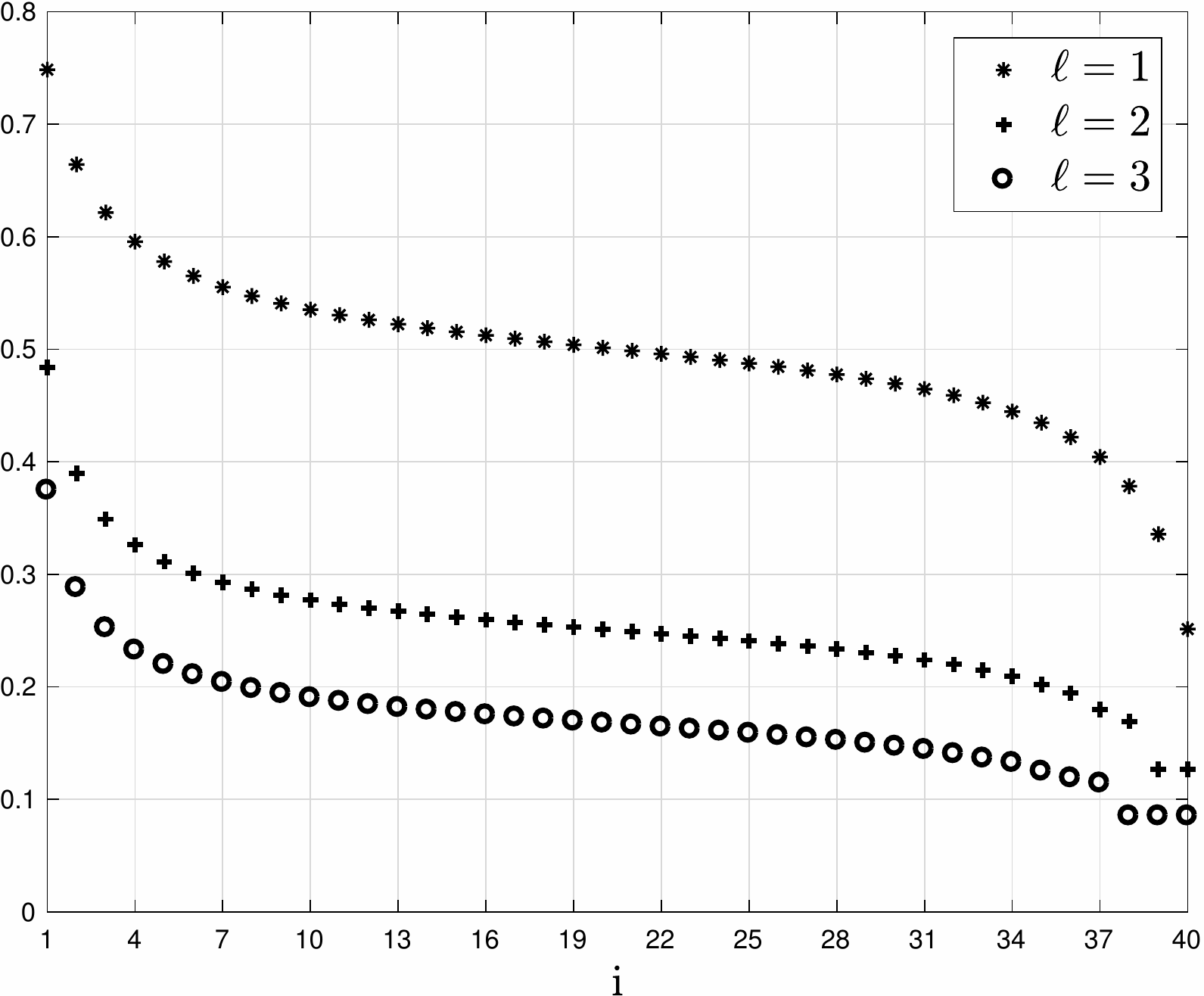}
\caption{Steady-state reader occupancy level $e_i$ as a function of $i=1,\dots,40$, for a TH{\model} with $n=40$ sites, and with $\ell=1$ (*), $\ell=2$ (+), and~$\ell=3$ (O).  }\label{fig:n40_l123_ss}
 \end{center}
\end{figure}

\begin{Example}
The steady-state reader occupancy levels of the {\model} with dimension~$n=40$ are depicted in Fig.~\ref{fig:n40_l123_ss} for three  particle sizes: $\ell=1$ (corresponding
 to the RFM), $\ell=2$, and~$\ell=3$. It may be observed that the steady-state reader occupancies monotonically decrease along the chain until the   last~$\ell$ densities that are equal.
The corresponding steady-state production rates are
 $R = 0.2513$ for~$\ell=1$;
  $R = 0.1265 $ for~$\ell=2$; and
  $R = 0.0851 $ for~$\ell=3$.~$\square$
\end{Example}

\subsubsection{Effect of particle size}
It is interesting to analyze how  the steady-state occupancies and production rate depend
on the particle size~$\ell$. One might naturally
expect the steady-state  production rate in the {\model} to decrease as the particle size~$\ell$
 increases.
Indeed, roughly speaking  one may think of increasing the particle size as
  replacing  small  cars traveling  along a unidirectional traffic lane with large trucks thus leading to
more congestion.
This is demonstrated by the next example.

\begin{figure}[t]
 \begin{center}
\includegraphics[width= 8cm,height=7cm]{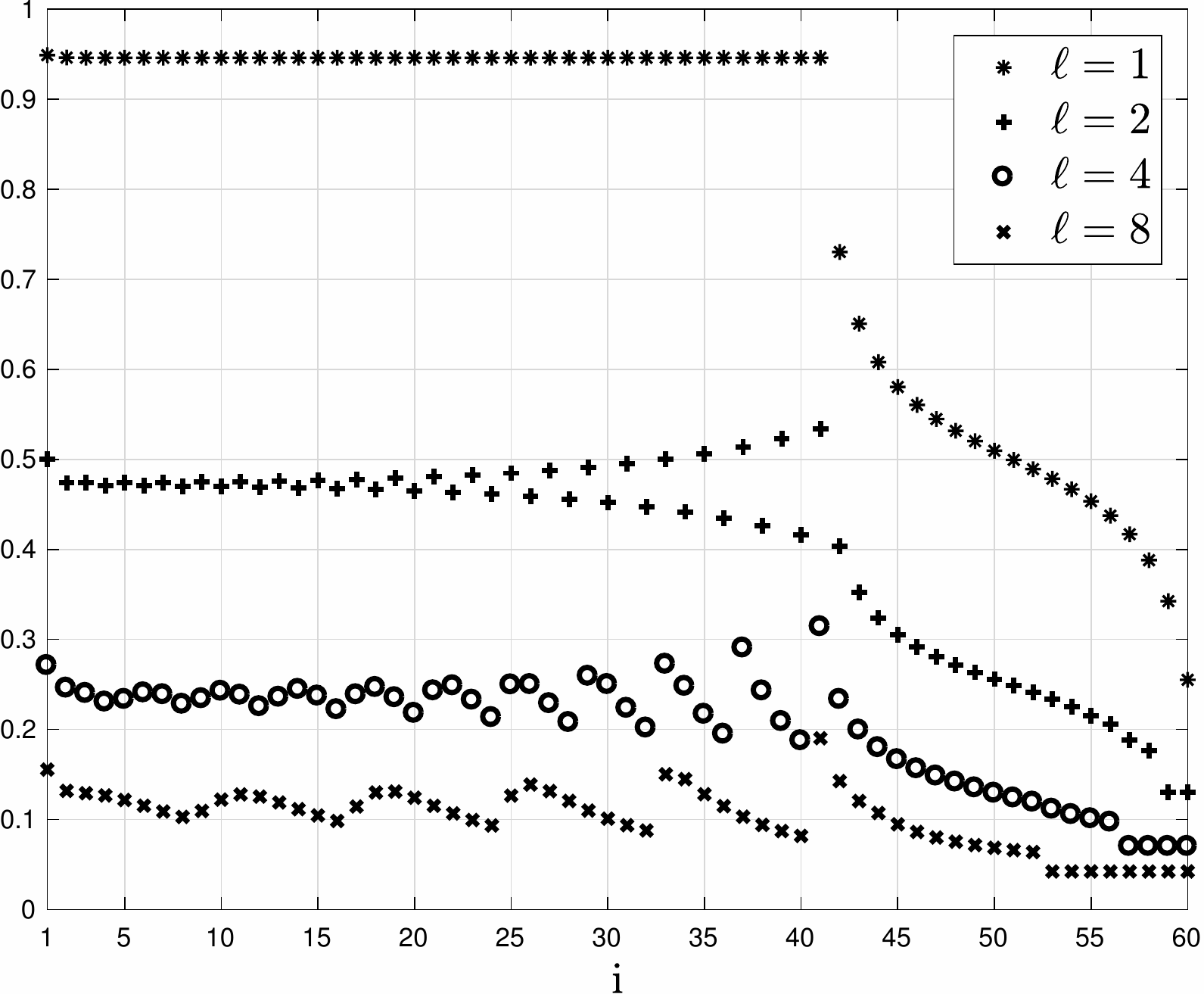}
\caption{Steady-state reader occupancy level $e_i$ as a function of $i=1,\dots,60$, for  a {\model} with $n=60$, $\lambda_0=\cdots=\lambda_{40}=1$, and $\lambda_{41}=\cdots=\lambda_{60}=1/5$, with $\ell=1$ (*), $\ell=2$ (+), $\ell=4$ (O), and $\ell=8$ (x).  }\label{fig:n60_l123_ss}
 \end{center}
\end{figure}

\begin{Example}
The steady-state reader occupancy levels in the
   {\model} with dimension~$n=60$, and rates~$\lambda_0=\cdots=\lambda_{40}=1$, and $\lambda_{41}=\cdots=\lambda_{60}=1/5$,  for four different
	particle sizes: $\ell=1$ (i.e. the RFM), $\ell=2$, $\ell=4$, and $\ell=8$
	are depicted in Fig.~\ref{fig:n60_l123_ss}.
Note that the steady-state occupancy levels decrease with $\ell$.
The transition rates here decrease from the value~$1$ to~$1/5$ at site~$40$. Thus, we except to see a ``traffic jam'' of ribosomes before this site. For~$\ell=1$ this is indeed what happens.
\updt{However, for~$\ell>1$ much more complicated patterns evolve.
 The steady-state densities follow a complicated quasi-periodic behavior, with period~$\ell$, even though there is
no such periodicity in the rates.}~$\square$
\end{Example}

\begin{figure}[t]
 \begin{center}
\includegraphics[width= 8cm,height=7cm]{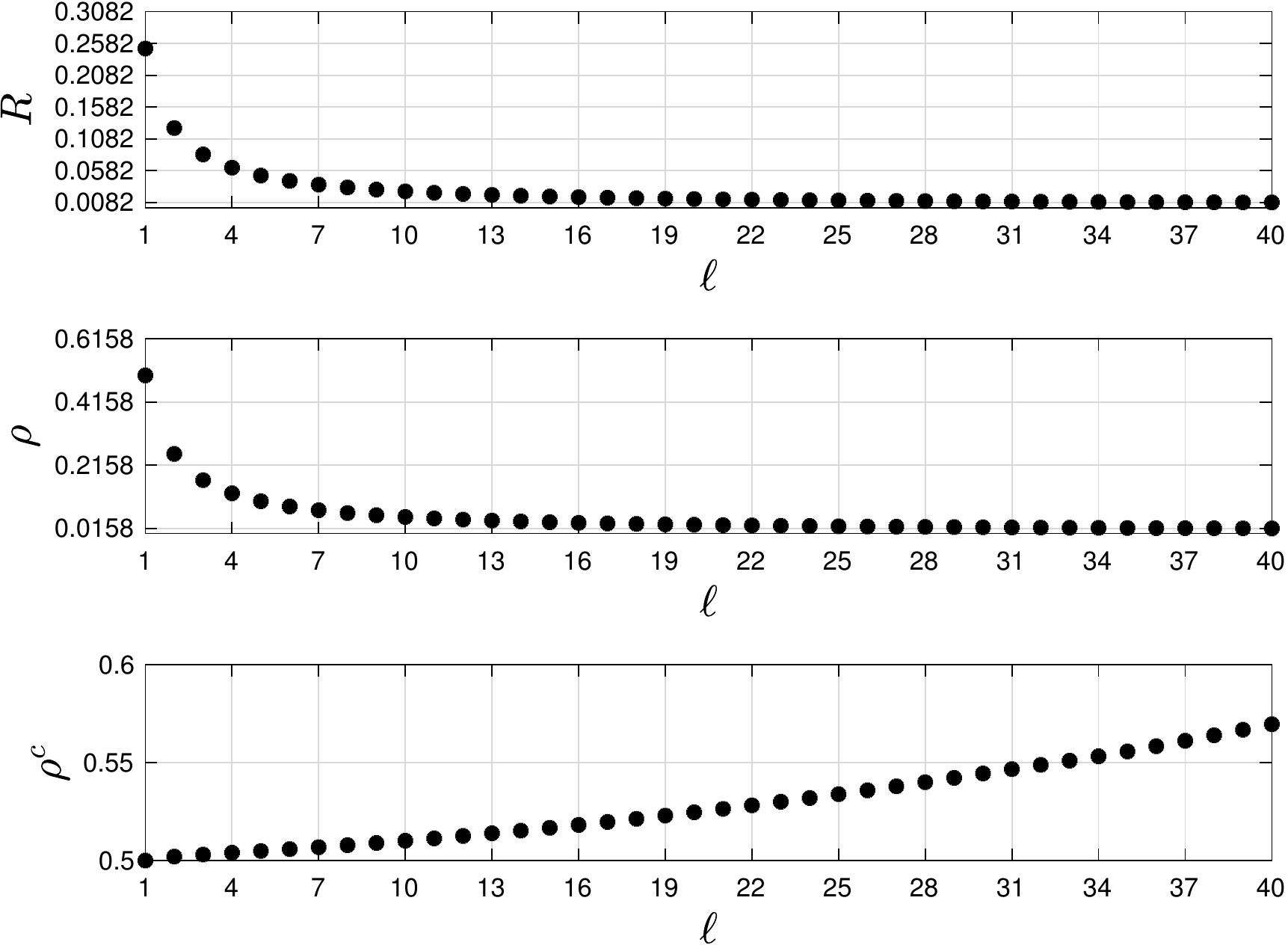}
\caption{Steady-state production rate $R$, mean reader occupancy $\rho$, and mean coverage occupancy $\rho^c$ as a function of $\ell$, for a TH{\model} with $n=100$ sites and $\lambda_q=1$. }\label{fig:n100_l_R}
 \end{center}
\end{figure}

\begin{figure}[t]
 \begin{center}
\includegraphics[width= 8cm,height=7cm]{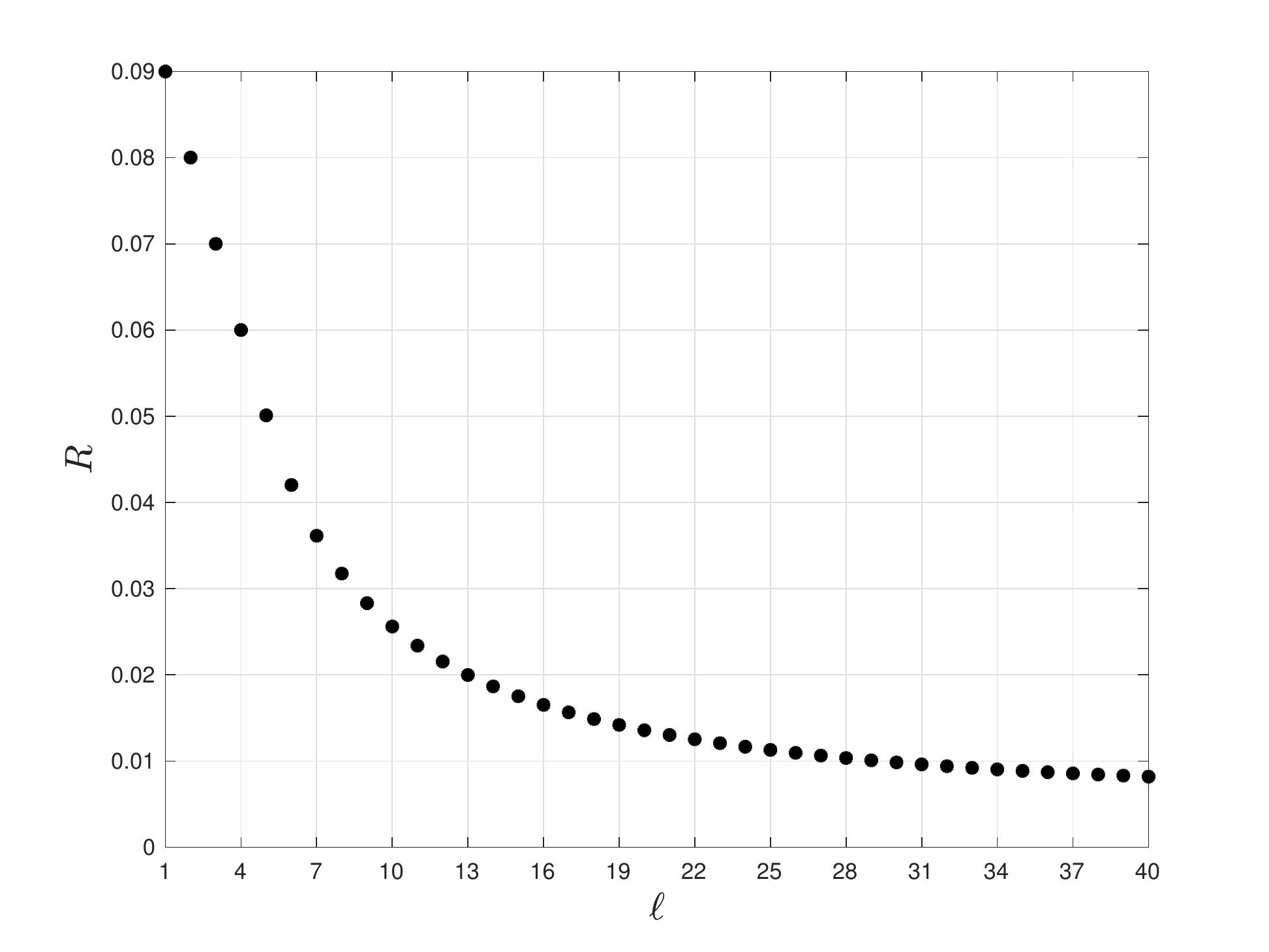}
\caption{\updt{The steady-state production rate $R$ as a function of $\ell$, for a {\model} with $n=100$, $\lambda_0=0.1$, and $\lambda_i=1$, $i=1,\dots,100$.} }\label{fig:n100_l_R_01}
 \end{center}
\end{figure}

\begin{Example}
Fig.~\ref{fig:n100_l_R} depicts the steady-state production rate $R$, the steady-state mean reader occupancy~$\rho$, and the steady-state mean coverage occupancy $\rho^c$ as a function of the particle size $\ell$, for a TH{\model} with dimension~$n=100$ and $\lambda_q=1$. It can be observed that the steady-state production rate and the mean reader occupancy decrease with $\ell$, whereas the steady-state mean coverage occupancy increases with $\ell$.

It is interesting to compare these results to the homogeneous TASEPEO.
In the thermodynamical limit (i.e. as~$N\to\infty$), the homogeneous TASEPEO with particle size~$\ell$, and with $\alpha=\beta=1$ is in the maximal current phase, where the steady-state output rate is~$J=1/(1+\sqrt{\ell})^2$,
 the mean  reader density is~$1/(\sqrt{\ell}(1+\sqrt{\ell}))$, and the mean  coverage density is~$  \sqrt{\ell}/(1+\sqrt{\ell)}$~\cite{shaw2004local,PhysRevE.76.051113}. Note that this implies that as~$\ell$ goes to infinity
the current and the mean reader density go to zero, whereas the mean coverage density goes to one. This \updt{is consistent with} the results for the TH{\model}
 depicted in Fig.~\ref{fig:n100_l_R}.

\updt{Fig.~\ref{fig:n100_l_R_01} depicts the steady-state production rate $R$ as a function of the particle size $\ell$ for a {\model} with $n=100$, $\lambda_0=0.1$, and $\lambda_i=1$, $i=1,\dots,100$. In this case $\lambda_0$ is the rate limiting factor, and thus less ``traffic jams" occur relative to the case $\lambda_0=1$. It may be seen that also in this case $R$ monotonically decreases with $\ell$.}~$\square$
\end{Example}

The next result shows that for fixed rates
the steady-state production rate in the {\model} with~$\ell>1$ is always smaller than the steady-state production rate in  the  {\model} with~$\ell=1$ (i.e. the~RFM).

\begin{Proposition}\label{prop:R_dec_l}
Consider an {\model} with dimension $n$, particle size $\ell>1$, and rates $\lambda_i$, $i=0,\dots,n$, admitting a steady-state production rate $R$. Consider also an~RFM with the same dimension $n$, and the
same rates $\lambda_i$, $i=0,\dots,n$, admitting a steady-state production rate $\bar R$. Then $R<\bar R $.
\end{Proposition}

 In many organisms longer genes have lower protein levels \cite{Dana2011,Eisenberg2003}. There are many explanations and variables that may contribute to  this correlation. However, is it possible that the relations between particle size and translation rate may have a (small) contribution to this correlation? It is possible that longer coding regions are related to longer proteins emerging from the ribosome during translation thus practically increasing the effective ribosome size. This hypothesis may be studied in synthetic system in the future.

Surprisingly, however, increasing~$\ell$ does not always lead to a reduction in the production rate.
\begin{Example}\label{exp:ll1}
Consider an {\model} with dimension $n=3$, and  rates
\[ \lambda_0=1.2, \;\lambda_1=0.8,\;\lambda_2=1.1,\;\lambda_3=3.
\]
We consider two cases~$\ell=2$ and~$\ell=3$, and for the sake of clarity we denote
 the steady-state values in the latter case by
overbars. For~$\ell=2$,
\[
e=\begin{bmatrix} 0.5213 & 0.2498 & 0.0916 \end{bmatrix}', \text{ and } R=0.2749.
\]
For~$\ell=3$,
\[
 \bar e=\begin{bmatrix} 0.3759 & 0.2733 & 0.1003 \end{bmatrix}',\text{ and } \bar R=0.3001.
\]
 \updt{Thus, here increasing~$\ell$ from~$2$ to~$3$ \emph{increases} the production rate.
To explain this, recall  that in
  general increasing~$\ell$   decreases   the steady-state reader densities (see Figs.~\ref{fig:n40_l123_ss} and~\ref{fig:n60_l123_ss}). This is also what happens here. Indeed,
\[
\bar e_1+\bar e_2+\bar e_3=0.7495 <   e_1+e_2=0.7711 .
\]
At steady-state, the entry rate into the chain is equal to the production rate, so
 $R=\lambda_0(1-e_1-e_2)$ and~$ \bar R=\lambda_0(1-\bar e_1-\bar e_2-\bar e_3)$.
Since this is proportional to one minus the sum of densities,~$\bar R> R$.
Thus, in this particular case the increase in~$\ell$ yields
  an \emph{increase} in the production rate.
	
	Similarly,  it follows from~\eqref{eq:rlm} and~\eqref{eq:rlm1} that for any~$n>3$
	increasing~$\ell$ from~$n-1$ to~$n$  in the TH{\model} leads to an increase in the steady-state production rate.}~$\square$
\end{Example}


\subsection{Entrainment}\label{subsec:entrain}
Assume now that some or all of the transition
rates~$\lambda_i$  are not constants, but
 time-varying \emph{periodic} functions with a common period~$T$.
In the context of translation, this corresponds for example to the case where the tRNA
abundances vary in a  periodic manner, with a common period~$T$.
More precisely, we  say that a function $f$ is $T$-periodic if $f(t+T) = f(t)$ for all $t$. Assume that the transition rates are  time-varying functions satisfying:
\begin{enumerate}
\item There exist $\delta_1,\delta_2\in\R_{++}$ such that $0<\delta_1\le\lambda_i(t)\le\delta_2$, for all~$i=0,\dots,n$, and  all $t\ge0$.
\item There exists a minimal $T>0$ such that every  $\lambda_i(t)$ is  a $T$-periodic function.
\end{enumerate}
We refer to this case as the \emph{periodic {\model} (P{\model})}. Note that the P{\model} includes in particular the case where some of the rates are constant, as a constant function is $T$-periodic for every $T$. However, condition~$2)$ above implies that the case where all the rates are constant is ruled out, as then the minimal~$T$
is zero. Indeed, this case is just the {\model} analyzed above.

The next result follows from combining the
 fact that the {\model} is~{\sostshort} on~$H$
with known results on the entrainment of contractive
systems to a  periodic excitation (see e.g.~\cite{entrain2011}).

\begin{Theorem}\label{thm:entrn}
The P{\model} admits a unique function $\phi(\cdot): \R_+ \to \Int(H)$, that is $T$-periodic, and for any~$a\in H $ the trajectory~$x(t,a)$ converges to~$\phi$ as~$t\to\infty$.
\end{Theorem}

In other words, the P{\model} admits a unique periodic solution, with period~$T$,
and every trajectory of the  P{\model} converges to this periodic solution.
This means that  the densities along the mRNA, and thus also the production rate entrain to the periodic excitation
induced by the transition rates.

As a side note, we point that the {\model} can also be used to model vehicular traffic.
If  traffic lights that change periodically produce the transition rates then
Thm.~\ref{thm:entrn} implies that the traffic density converges to  a periodic pattern with the same period, i.e.
the ``green wave'' concept  (see, e.g.,~\cite{gwave2013}).

The next example illustrates the dynamical behavior of the P{\model}.

\begin{Example}\label{exp:per}
Consider an P{\model} with dimension~$n=4$, particle size~$\ell=2$, and transition rates
 \begin{align*}
\lambda_0(t)&=1+0.5\sin(\pi t/4),\\
\lambda_1(t)&\equiv0.9,\\
\lambda_2(t)&\equiv0.8,\\
\lambda_3(t)&=1+0.25\sin(  (1+\pi t )/2   ),\\
\lambda_4(t)&\equiv1.4.
\end{align*}
 Note that all the rates here are periodic, with a minimal common period $T = 8$. Figure~\ref{fig:rfmeo_per} depicts~$x_i(t)$,
$i = 1, \dots, 4$, as a function of $t$
for the  initial condition $x(0)=\begin{bmatrix} 0.2 & 0.2 & 0.2 & 0.2 \end{bmatrix}'$.
 It may be seen that each state-variable converges to a periodic function with period $T = 8$.~$\square$

\begin{figure}[t]
 \begin{center}
\includegraphics[width= 8cm,height=7cm]{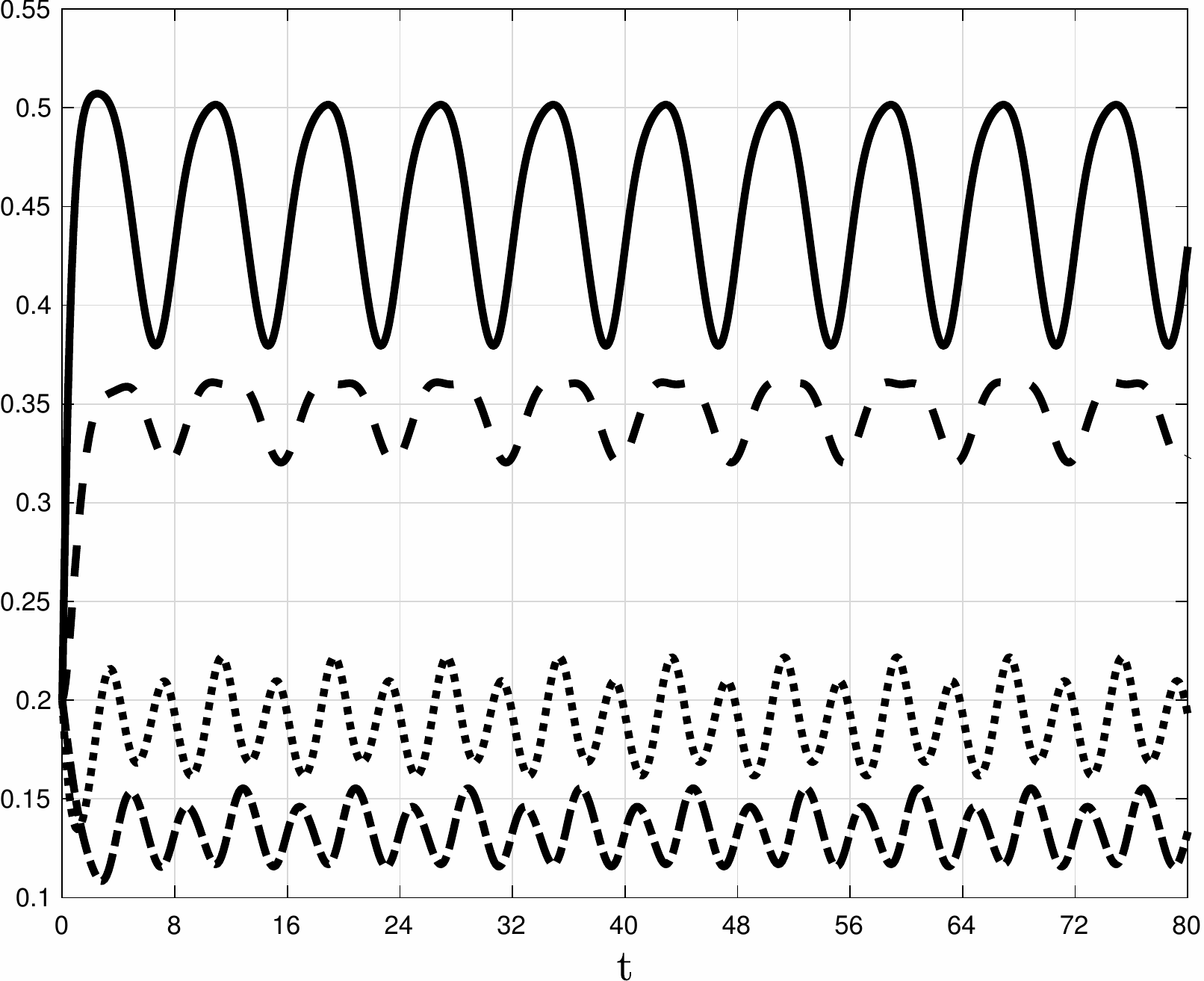}
\caption{State variables $x_1(t)$ [solid line]; $x_2(t)$ [dashed line]; $x_3(t)$ [dotted line]; and $x_4(t)$ [dashdot line] as a function of $t$ in Example~\ref{exp:per}. Note that each state-variable converges to a periodic function with  period $T = 8$. }\label{fig:rfmeo_per}
 \end{center}
\end{figure}
\end{Example}

\subsection{Rate limiting steps in the RFM and the {\model}}
It has been shown that depending on the biological
conditions and the specific  organism both initiation and elongation may be rate limiting \cite{Tuller2015,Ciandrini2013,Zur2016,Racle2013,Tuller2010,Haar2012,Jacques1990}.
Since the {\model} is a better model for biological translation than the RFM, it is interesting to
study the rate limiting step in these two models.
We now show that the transition from the low density phase, when initiation is rate limiting, to the high density phase, when elongation is rate limiting is different in the two models:
in the {\model}
 this transition will take place    for
a lower initiation rate.

We modeled four  \emph{S.~cerevisiae} genes: YMR123W, YNL303W, YJR094W-A, and YBL094C using
 both an {\model} with~$\ell=10$  and an RFM,  and  considered  the steady-state production rate and the steady-state mean density as a function of the initiation rate $\lambda_0$.

\updt{As was done in Example~\ref{exp:scrv_dist},} the  elongation rate~$\lambda_i$  at  each site, for both the {\model} and the RFM, was estimated  using ribo-seq data for the codon decoding rates 
 normalized so that the median elongation  rate of all {\em S. cerevisiae} mRNAs
becomes~$6.4$ codons per second. 
The site rate is simply the corresponding codon rate.
These rates thus  depend on
various factors including  availability of tRNA molecules, amino acids, Aminoacyl tRNA synthetase activity and concentration, and local mRNA folding~\cite{Dana2014B,Alberts2002,Tuller2015}.

\begin{figure}[t]
 \begin{center}
\includegraphics[width= 9cm,height=8cm]{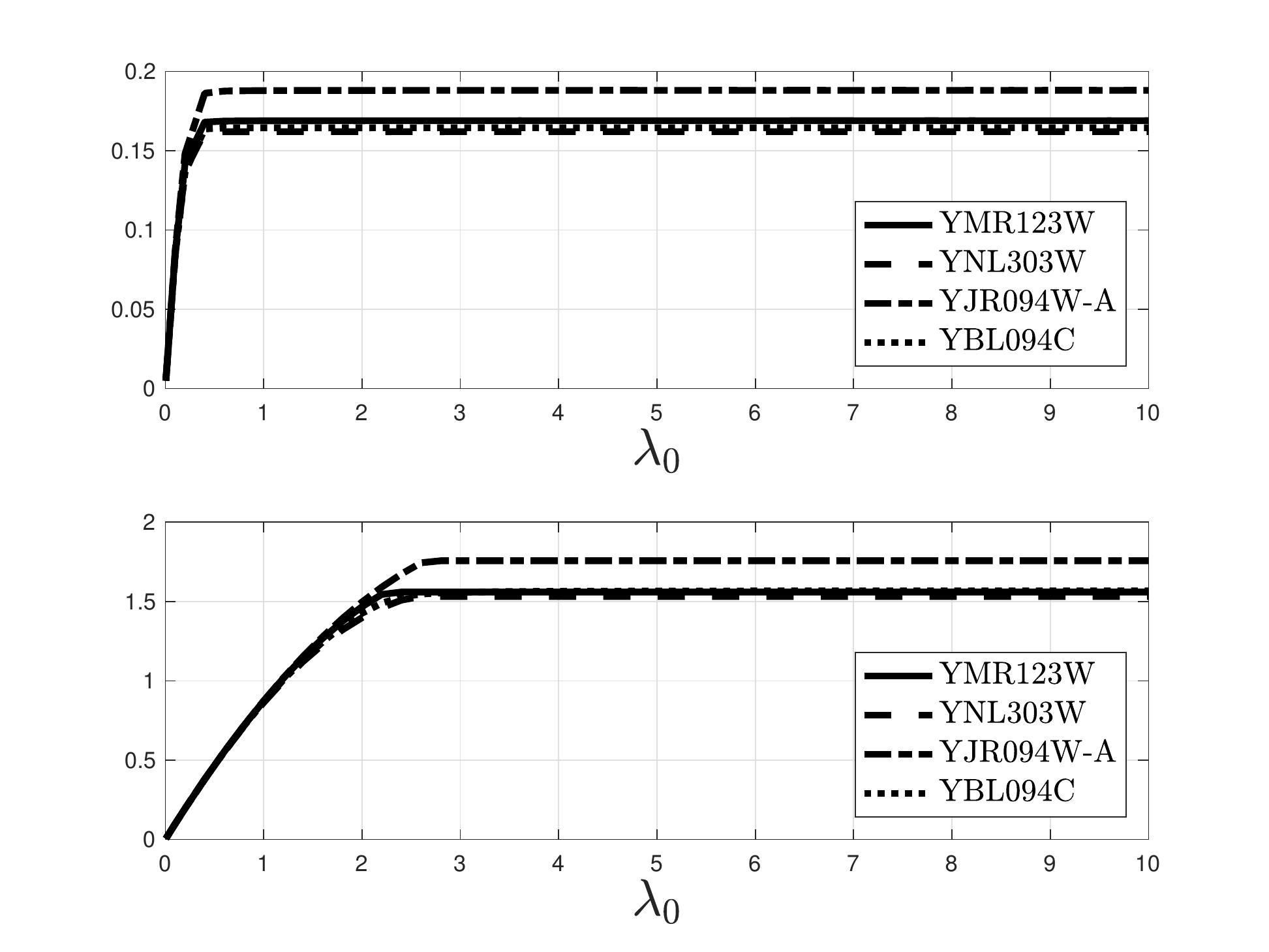}
\caption{The steady-state production rate $R$ as a function of $\lambda_0$ for four
 \emph{S. cerevisiae} genes. Upper figure: {\model}. Lower figure: RFM.  }\label{fig:R_l0_rfm_rfmeo}
 \end{center}
\end{figure}

Fig.~\ref{fig:R_l0_rfm_rfmeo} depicts the steady-state production rate as a function of $\lambda_0$ for the four \emph{S. cerevisiae} genes for both the {\model} (upper figure) and the RFM (lower figure). It may be seen that the transition from an initiation rate limiting stage to an elongation rate limiting stage occurs for a  lower initiation value in the~{\model} as compared to the~RFM.

\begin{figure}[t]
 \begin{center}
\includegraphics[width= 9cm,height=8cm]{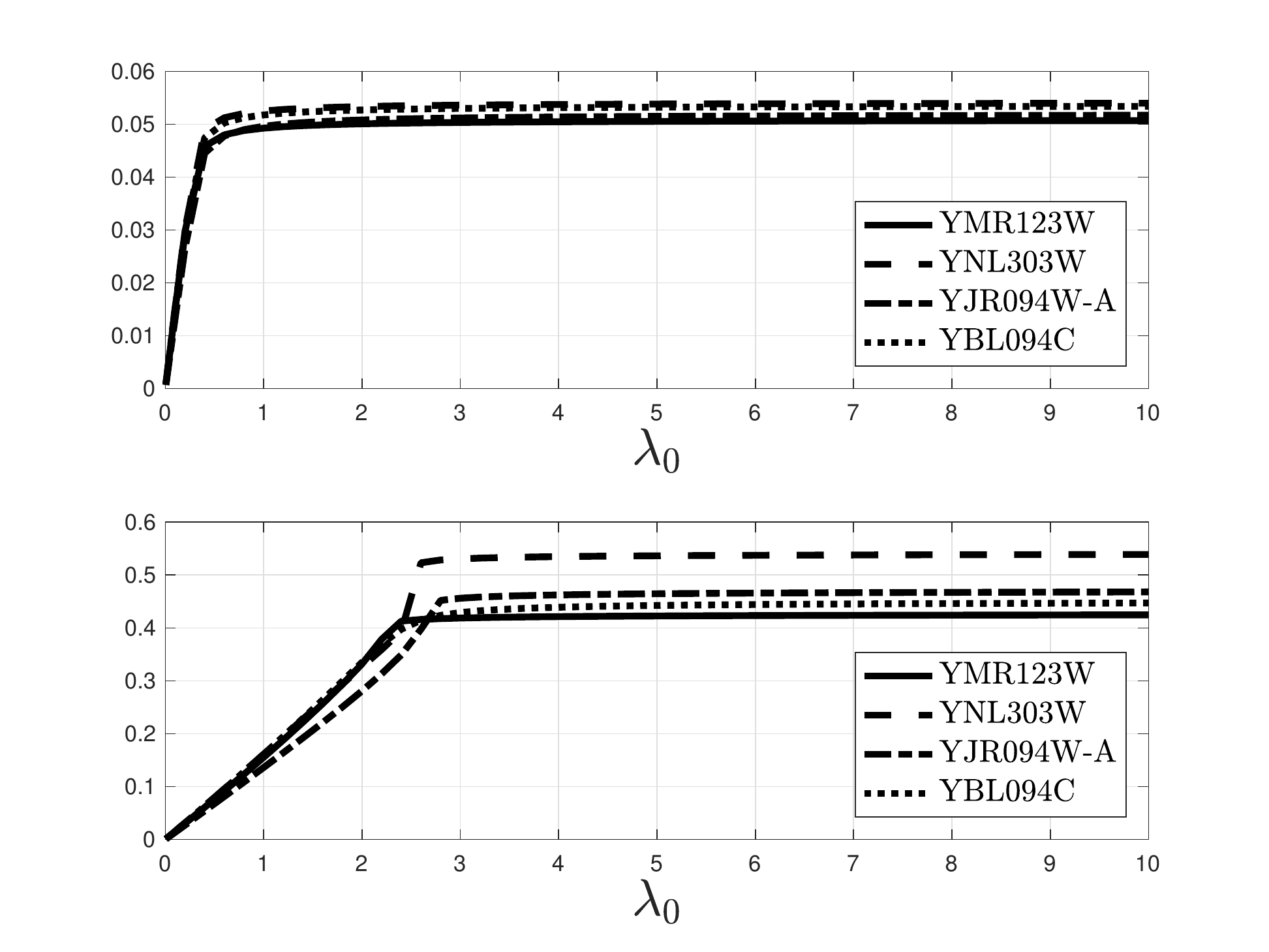}
\caption{The steady-state mean density $\rho$ as a function of $\lambda_0$ for four
\emph{S. cerevisiae} genes. Upper figure: {\model}. Lower figure: RFM.   }\label{fig:rho_l0_rfm_rfmeo}
 \end{center}
\end{figure}

Fig.~\ref{fig:rho_l0_rfm_rfmeo} depicts the steady-state mean density as a function of $\lambda_0$ for the four
  genes and two models. Again, it   can be seen
  that the transition from an initiation rate limiting stage to the  elongation rate limiting stage occurs at lower initiations value in the~{\model} as compared to the~RFM. This holds for all four genes.

\section{High correlation between {\model} and TASEPEO}\label{sec:compare}
In this section, we show that
 the {\model} correlates better with
   TASEPEO   than the RFM, supporting the modeling
	of intracellular process with multi-site biological machines
	such as translation and transcription using  the {\model}.

The simulations of~TASEPEO with dimension $N$, rates $\mu$ (see~\eqref{eq:tasep_rates}), and particle size $\ell$ use a parallel update mode.
 At each time tick~$t_k$, the sites along the lattice are scanned from site~$N$ backwards to site~$1$.
 If it is time to hop,  and the site that is $\ell$ sites in front is empty then  the reader advances to the consecutive site.
If  the site that is $\ell$ sites in front is occupied, the next hopping time,~$t_k+\varepsilon_k$,
is generated randomly. For site~$i$, $\varepsilon_k$
 is  exponentially distributed with parameter~$(1/\mu_{i+1})$ (see~\eqref{eq:tasep_rates}). The occupancy at each site is averaged throughout the simulation,
with the first~$700,000$ cycles discarded in order to obtain the steady-state value.
We use~$\varrho\in\R^N_{+}$ to denote the steady-state reader density,
$J:=\beta \varrho_N$ to denote  the steady-state current (or output rate), and $\sigma:=(1/N)\sum_{i=1}^N \varrho_i$ for the steady-state mean reader density.

In the examples below, we  numerically calculated
 the   Pearson correlation coefficients between the steady-states of the {\model}, TASEPEO, and RFM.

\begin{Example}\label{exp:synth_eoz}
Consider the {\model} with dimension $n=75$, and transition rates~$\lambda_0=\cdots=\lambda_{75}=1$. Let $\tilde e$ denote the steady-state density of an RFM with the same dimension  and rates. We also simulated~TASEPEO with dimension $N=75$ and rates~$\mu=\lambda$.
Fig.~\ref{fig:comp_synth_eoz} depicts the Pearson correlation coefficient $r(e,\varrho)$ between the steady-state reader densities of the {\model} and the TASEPEO, and the Pearson correlation coefficient~$r(\tilde e, \varrho)$ between
 the steady-state reader densities of the RFM and TASEPEO, as a function of~$\ell\in\{1,\dots,30\}$. The corresponding p-values were all less than~$10^{-50}$.
It may be seen that~$r(e,\varrho)$ and~$r(\tilde e, \varrho)$ are somewhat similar for~$\ell\in\{1,\dots,5\}$, however for all~$\ell>5$, $r(e,\varrho)>0.94$ whereas~$r(\tilde e,\varrho)$ decreases with~$\ell$, and is equal to about~$0.825$ for~$\ell=30$. Of course, this makes sense as the~{\model} is a  mean field
approximation  of~TASEPO.~$\square$

\begin{figure}[t]
 \begin{center}
\includegraphics[width= 8cm,height=7cm]{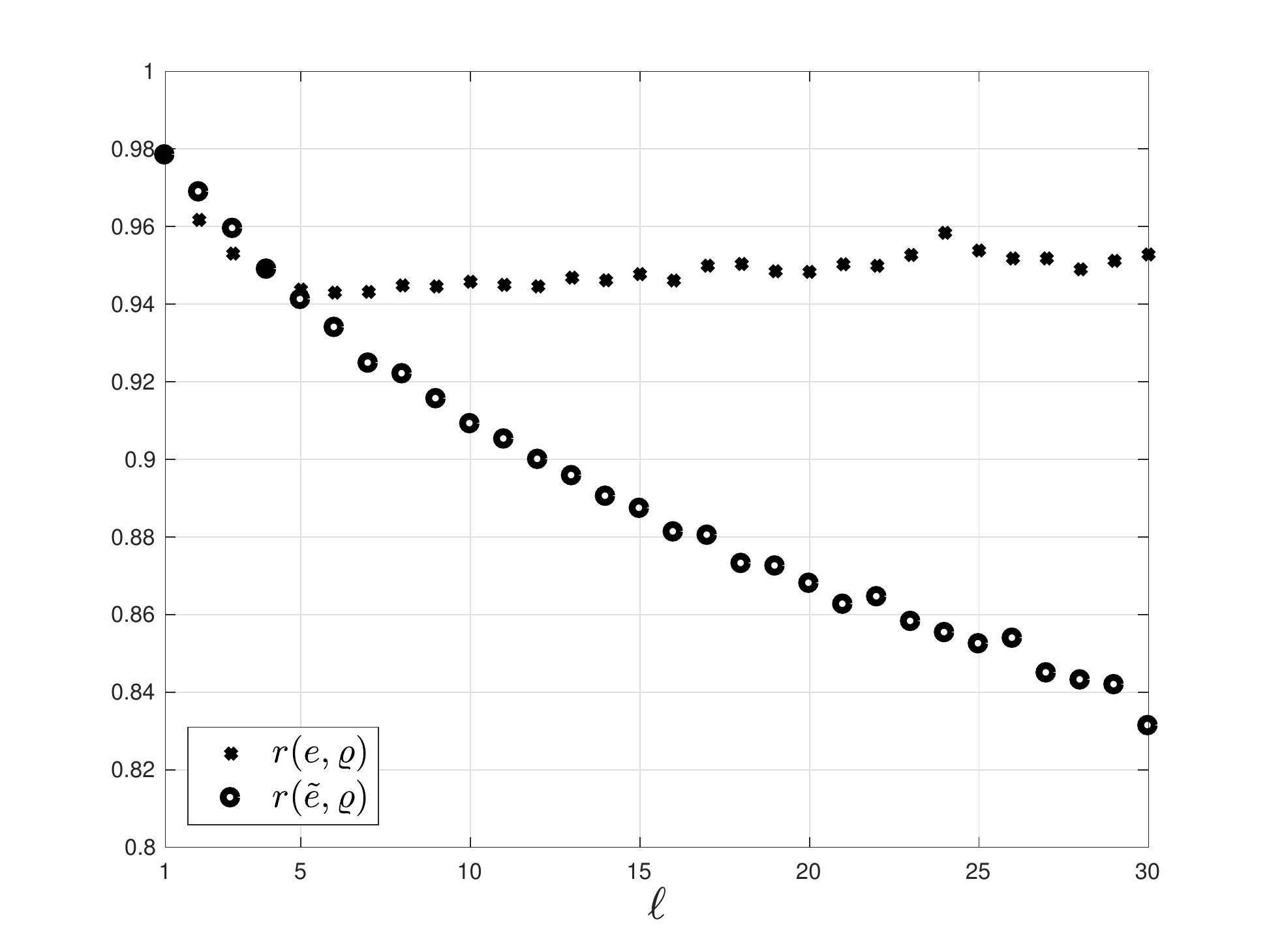}
\caption{Pearson correlation coefficient $r$ between the steady-state reader densities of the {\model} ($e$) and TASEPEO ($\varrho$), and between the steady-state reader densities of the RFM ($\tilde e$) and TASEPEO ($\varrho$), for $\ell\in\{1,\dots,30\}$. }\label{fig:comp_synth_eoz}
 \end{center}
\end{figure}

\end{Example}

\begin{figure}[t]
 \begin{center}
\includegraphics[width= 8cm,height=7cm]{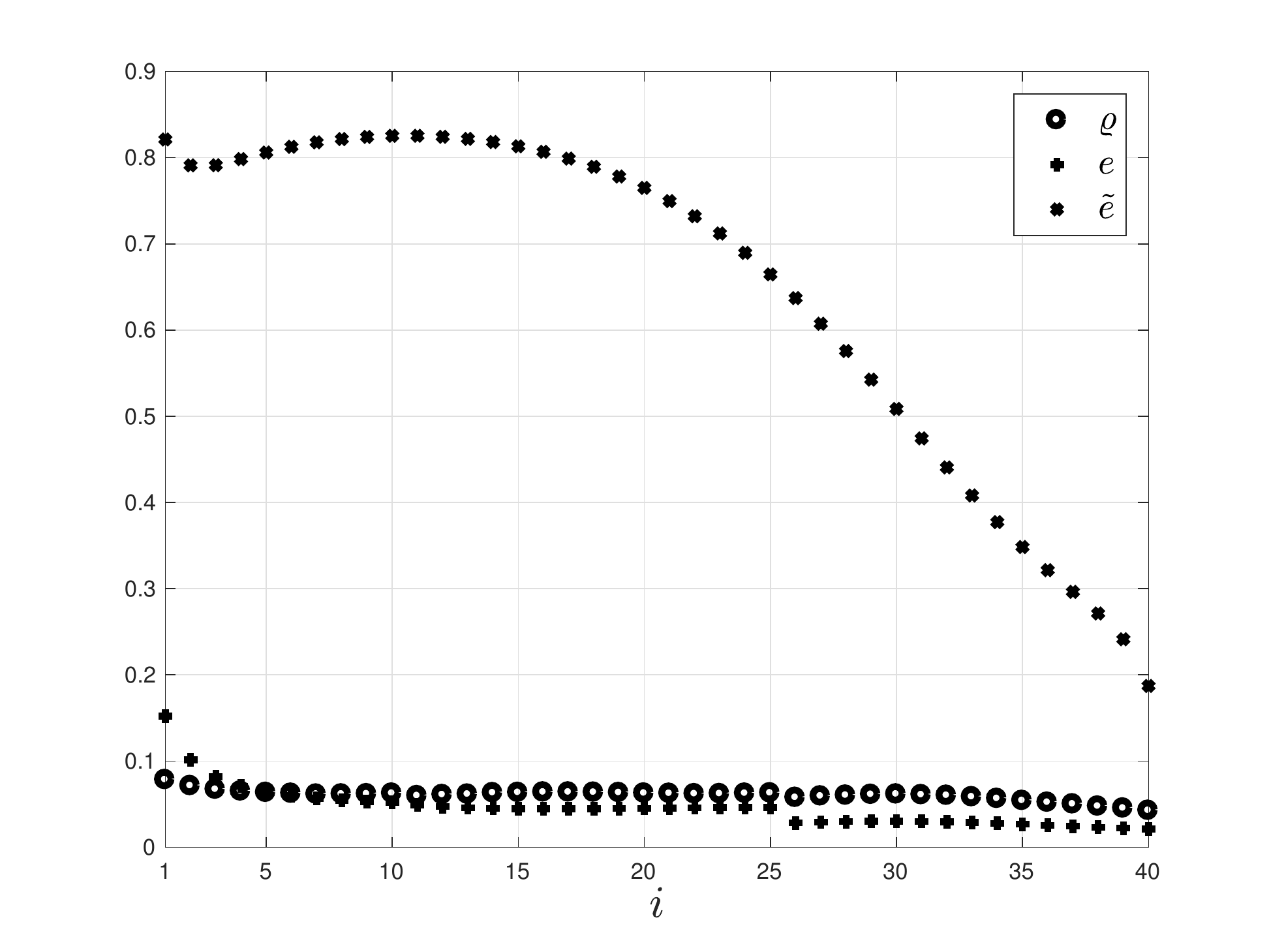}
\caption{Steady-state density as a function of the site number~$i$
 for  the {\model} ('+'),  TASEPEO ('o'), and  RFM ('*')
in Example~\ref{exp:denst_n40}. }\label{fig:denst_n40}
 \end{center}
\end{figure}

The following examples consider the case of non-homogeneous transition rates.
\begin{Example}\label{exp:denst_n40}
Consider the {\model} with dimension $n=40$, particle size $\ell=15$, and rates $\lambda_i=1+0.3\sin(2\pi i/41)$, $i=0,\dots,40$. Fig.~\ref{fig:denst_n40} depicts the {\model} steady-state reader density $e$, the TASEPEO steady-state reader density $\varrho$ for $\mu=\lambda$, and particle size $15$, and the  steady-state density $\tilde e$ in the~RFM
with the same dimension and rates. It can be seen that
 $e$ provides a far  better estimate of~$\varrho$ than~$\tilde e$.~$\square$
\end{Example}

In order to verify that the high correlation between {\model} and TASEPO
holds for a large set of parameters, we also simulated
 the case where the rates are drawn randomly.

\begin{Example}\label{exp:R_rho_corr}
Consider the {\model} with dimension $n=100$, particle size $\ell=10$, and rates
\be\label{eq:R_rho_corr}
\lambda_i=1+\theta_i,  \quad i=0,\dots,100,
\ee
where $\theta_i\sim U[-1/2,1/2]$ is a random variable uniformly distributed in the interval $[-1/2, 1/2]$.
We compared the steady-state production rates of this {\model} with those of the corresponding TASEPO, and with two
RFMs. One  RFM with the same dimension and rates.
Another RFM, that we refer to as~RFM10, is an approximation of the chain
with $10$ ``codons''/site.
Thus, it has dimension
 $(100/10)-1=9$,
  where each site contains $10$ consecutive sites of the {\model} (other than the last site which contains the last $11$ consecutive sites of the {\model}). The rates of RFM10 are
	$\gamma_i=(\sum_{k=10 i}^{T_i} \lambda_k^{-1})^{-1}$, where $T_i=(10(i+1)-1)$ if $i<9$, and otherwise~$T_i=100$.  Note that since the dimension of this
	RFM$10$   is nine, it cannot be used to estimate the entire
	density profile of the TASEPEO with dimension $100$.

We ran $300$ tests, where in each test a new set of rates
were drawn
 according to~\eqref{eq:R_rho_corr}. Fig.~\ref{fig:R_corr} depicts the correlation between the steady-state production rates of \updt{(1)~{\model} and TASEPEO;
(2)~RFM (i.e. RFM with dimension $100$ and rates $\lambda_i$) and~TASEPEO, and
(3)~RFM$10$ and TASEPEO}, over the $300$ tests. It may be seen that the~{\model} provides the best correlation with  TASEPEO.

\begin{figure}[t]
 \begin{center}
\includegraphics[width= 8cm,height=7cm]{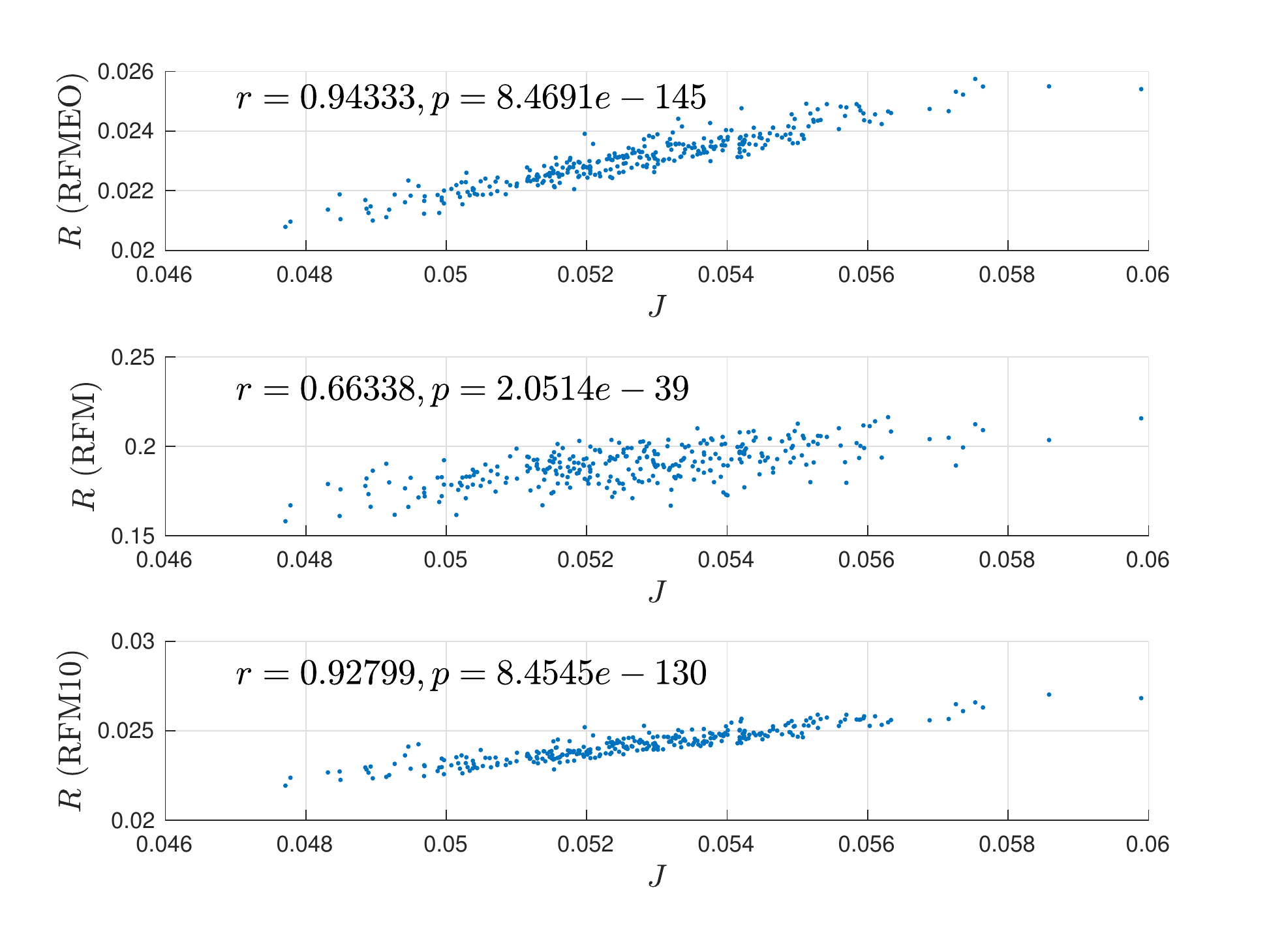}
\caption{Steady-state production rates, and the corresponding Pearson correlation
coefficient $r$  and  $p$-value in Example~\ref{exp:R_rho_corr}. Upper: Steady-state production rates~$R$ in
 {\model} vs.  $J$ in TASEPEO; Middle: Steady-state production rates $R$
in the RFM  vs. $J$ in TASEPEO;
Lower: Steady-state production rates  $R$ in  RFM$10$ vs.  $J$ in TASEPEO. }\label{fig:R_corr}
 \end{center}
\end{figure}

Fig.~\ref{fig:rho_corr} depicts the correlations between the steady-state mean densities for the same three cases.  It may be seen that again the correlation between the {\model} and TASEPEO is high ($r\simeq 0.927$). The correlation between the RFM$10$ and  TASEPEO is slightly better ($r\simeq0.944$), however, as stated above,  RFM10 cannot be used to provide an estimate to the actual (per codon) density profile.~$\square$

\begin{figure}[t]
 \begin{center}
\includegraphics[width= 8cm,height=7cm]{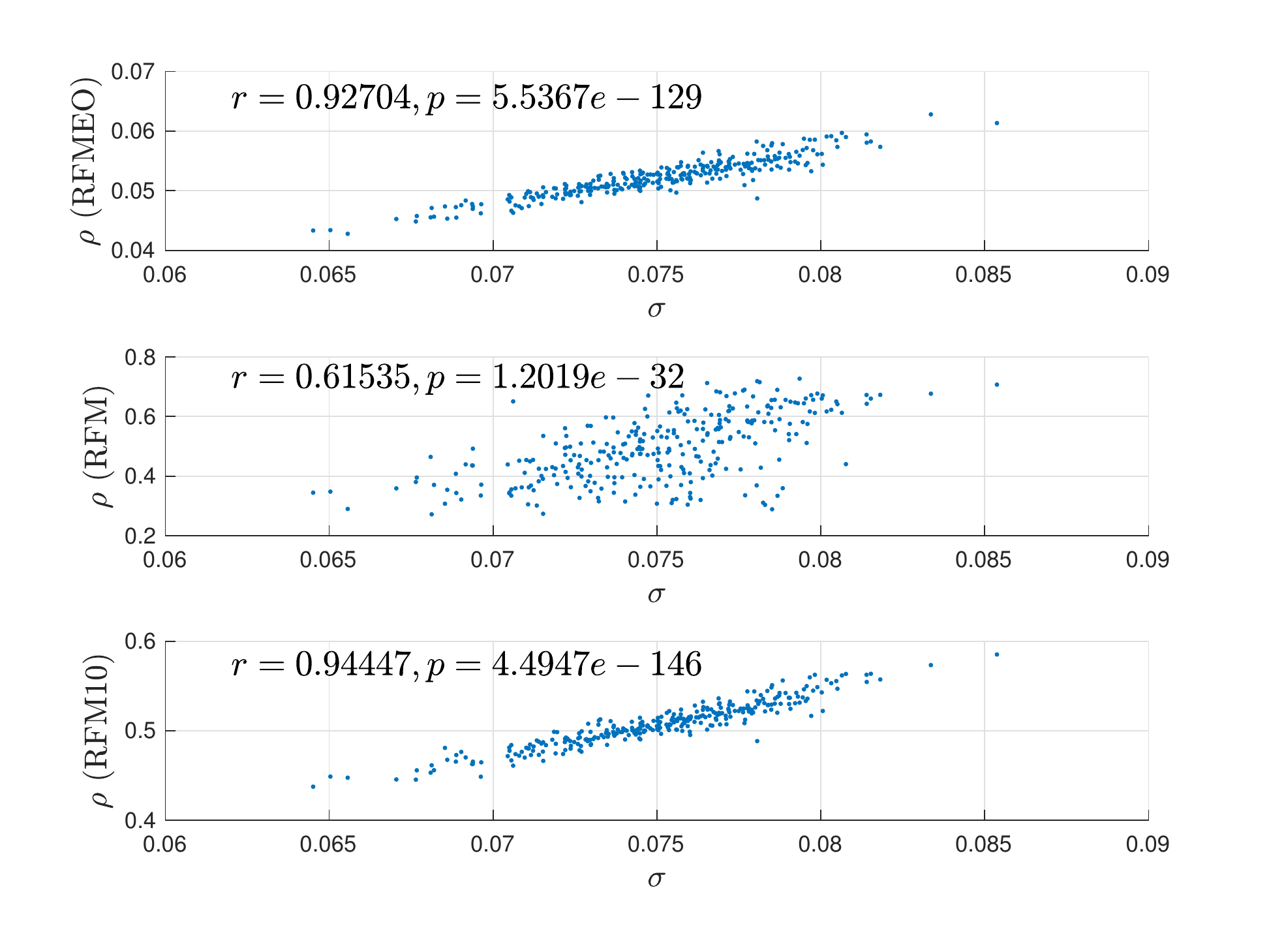}
\caption{Steady-state mean densities, and the corresponding Pearson correlation coefficient $r$ value and
 $p$-value in Example~\ref{exp:R_rho_corr}. Upper: Steady-state mean density $\rho$ in the {\model}
vs. $\sigma$ in TASEPEO; Middle: Steady-state mean density $\rho$
in the RFM vs. $\sigma$ in TASEPEO;
Lower: Steady-state mean density~$\rho$
 in the RFM$10$  vs. $\sigma$ in TASEPEO. }\label{fig:rho_corr}
 \end{center}
\end{figure}

\end{Example}

\section{Discussion}

We studied a deterministic mechanistic model for mRNA translation, the {\model},
 that
 encapsulates many realistic features of this biological process including the fact that every
ribosome covers several codons and that   ribosomes cannot overtake one another.

The {\model} is a  mean-field approximation of~TASEPEO (see Appendix~B) and, as demonstrated above,
 its simulation results
 often
correlate well with  those of~TASEPEO. However, unlike TASEPEO, the~{\model}
   is amenable to rigorous analysis using tools from systems and control theory.

\updt{We proved  that the {\model} converges to a unique state-state density and steady-state production rate for any set of feasible transition rates. \updt{We  follow the  terminology used in physics, where an equilibrium point [steady-state] is characterized by a zero [constant but nonzero] total  flow of energy~\cite{STEADY_IN_BIO}}.
The convergence to this unique steady-state takes place at an  exponential rate.
In this respect, the {\model} is robust to the initial conditions.

One  may naturally ask whether biological systems are at steady-state (that maybe more general than the steady-state here, e.g. a
periodic trajectory).  Models with a steady-state (or several  of steady-states)
have been found to be useful in numerous studies in systems biology
 (see, e.g.~\cite{sys_bio_book2010} and the references therein). 
 In practice the state of the art routine biological experiments and their 
 interpretation assume steady state as they are performed in a very specific
experimental environment which is kept constant during the entire experiment (see, for example, \cite{Mukherji2011,Vogel2012,Bar-Joseph2012}).
 
  In particular, the steady-state in the~RFM
has been used
  to   accurately predict several features of gene expression (see, e.g., \cite{reuveni,Zur2016,Edri2014}).
Here, we used the {\model} to
model a
 highly-expressed \emph{S. cerevisiae} gene. The rates were estimated  based on biological data.
In the resulting {\model} the convergence to a state close to the steady-state takes approximately~30 seconds,
whereas the mRNA half-life is of the order of tens of minutes. This suggests that at least in this case the
steady-state assumption is justified.  }

 \updt{An important question is how does the steady-state depend on the {\model} parameters.  We proved
 that increasing any of the {\model} rates can only increase the steady-state production rate, and that in the totally homogeneous case (i.e. when all the rates are equal) the reader ribosomal density monotonically decreases along the mRNA. In addition, we proved  that if one or more of the {\model} rates are time-varying periodic functions, with a common period~$T$, then the densities along the mRNA, and thus also the production rate converge to a periodic solution with period~$T$. }

The results reported here can shed  light on various biophysical aspects of translation, and can be further studied experimentally. For example, our analysis suggests that  higher decoding rates at the last $\ell$ codons of the coding region can be expected (since in this region no downstream ribosome can block the ribosome movement). This can be validated experimentally for example based on approaches that track the movement of ribosomes at high resolution~\cite{Uemura2010}.

In addition,  analysis and simulations of the {\model} demonstrate several surprising and counterintuitive results.
For example, increasing the particle size~$\ell$ (i.e. the ribosome footprint) may some times lead to an increase in the production rate. Also, for large~$\ell$ the steady-state density along the mRNA
may be quite complex (e.g. with quasi-periodic patterns) even for relatively simple (and non-periodic)
transition rates. It will be interesting to see if similar patterns are observed experimentally by possibly engineering the codon elongation rates of heterologous or endogenous genes and monitoring translation \cite{Uemura2010,Ingolia2009}.

We believe that the {\model}
could  be   useful   for modeling, understanding, and re-engineering translation. Specifically, the advantages of the model mentioned above should make it a better candidate than other
 alternative models for solving some of the open questions in the field~\cite{Zur2016}.

\updt{
 An important topic for future research is using the {\model}
to model ribosome flow based on biological data. This is a
challenging task,  as many aspects of translation are still not   clear.
For example,  translation initiation is affected by complex  phenomena  such
 as the number of free ribosomes,   mRNA folding near the 5'end of the mRNA,   UTR length and other features, the nucleotide composition surrounding the start codon, and more.
In addition, current techniques for measuring
  ribosome densities  provide    partial, noisy, and   biased data (see, for example,~\cite{Diament2016}).
	Thus, using them
	to estimate  the parameters in a computational  model like the {\model}
	 is a non trivial challenge.}

Another  research topic  is  using the {\model} (and  networks  of~{\model}s)
 to study various phenomena such as competition  for  resources in mRNA translation~\cite{Raveh2016,Zur2016}, transcription \cite{edri2013}, and evolution of transcripts~\cite{Zur2016}.

\section*{Acknowledgments}
\updt{We are grateful to the anonymous referees for their comments that greatly helped in improving this paper. }

\section*{Appendix A: Proofs}

{\sl Proof of Prop.~\ref{prop:ydun}.}
Combining~\eqref{eq:y} and~\eqref{eq:rfmeo_ode} yields
\begin{align}\label{eq:rfmeo_ode_yx}
\dot{y}_i=\sum_{m=1}^i \dot{x}_m=&\lambda_{0}(1-y_{\ell})-\lambda_i x_i (1-y_{i+\ell}), &&\quad 1\le i\le\ell,  \nonumber \\
\dot{y}_i=\sum_{\mathclap{m=i-\ell+1}}^i \dot{x}_m=&\lambda_{i-\ell}  x_{i-\ell}  (1-y_{i})- \lambda_i x_i (1-y_{i+\ell}), &&\quad  \ell<i\le n.
\end{align}
By the definition of~$y_i$,
$
x_i=y_i-y_{i-1}+x_{i-\ell},
$
and iterating this yields
\be\label{eq:rfmeo_ode_xy}
x_i=\sum_{k=0}^{\lceil (i-\ell)/\ell \rceil} (y_{i-k\ell}-y_{i-k\ell-1}).
\ee
Substituting  this in~\eqref{eq:rfmeo_ode_yx} yields~\eqref{eq:rfmeo_ode_y}.~\IEEEQED

{\sl Proof of Prop.~\ref{prop:repel}.}
Consider the {\model} with~$x(0)\in \partial H$.
 Then~$y(0)=Px(0)$, and there exists an index~$i$ such that
either~$x_i(0)\in\{0,1\}$ or~$y_i(0)\in\{0,1\}$ and all the other entries of~$x(0)$ and~$y(0)$ are between zero and one.
  The proof is based on computing the derivatives of the state-variables at time zero, and showing
that state-variables that are zero [one] become strictly larger than zero [strictly smaller than one] at time~$0^+$.
We assume throughout that~$\ell\geq 2$, as otherwise the {\model} reduces to the RFM and then the proof follows from the results
in~\cite{RFM_entrain}.
 We consider several cases.

\noindent Case 1.
Suppose that~$y_\ell(0)=0$. This implies in particular  that~$x_\ell(0)=0$.
By~\eqref{eq:rfmeo_ode_yx},
\begin{align*}
					\dot y_\ell (0) &= \lambda_{0} (1-y_\ell (0))-\lambda_\ell x_\ell (0) (1-y_{2\ell} (0))\\
					         &= \lambda_0.
\end{align*}
Thus,~$y_\ell(0^+)>0$. Note that this calculation
also implies that for any~$\tau>0$ there exists~$\varepsilon_\ell=\varepsilon_\ell(\tau)>0$ such
that~$y_\ell(t,a)\geq \varepsilon_\ell  $ for all~$t\geq\tau$ and   all~$a\in H$.

\noindent Case 2.
Suppose that~$y_{\ell+1}(0)=0$. This implies in particular  that~$x_{\ell+1}(0)=0$,
so~$ y_\ell(0)=y_\ell(0)-y_{\ell+1}(0) =x_1(0)-x_{\ell+1}(0)=x_1(0)$.
By~\eqref{eq:rfmeo_ode_yx},
\begin{align*}
					\dot y_{\ell+1}(0)&= \lambda_{1} x_1 (0)(1-y_{\ell+1}(0))-\lambda_{\ell+1} x_{\ell+1} (0)
					(1-y_{2\ell+1}(0))\\
					         &= \lambda_1   y_\ell(0) .
\end{align*}
Combining this with the result in  Case~1  implies  that for any~$\tau>0$ there exists~$\varepsilon_{\ell+1}=\varepsilon_{\ell+1}(\tau)>0$ such
that~$ y_{\ell+1}(t,a)\geq \varepsilon_{\ell+1}  $ for all~$t\geq\tau$ and   all~$a\in H$.

Continuing in this fashion shows that for
 any~$\tau>0$ there exists~$\varepsilon=\varepsilon(\tau)>0$ such
that~$ y_i(t,a)\geq \varepsilon   $ for all~$i\in\{\ell,\ell+1,\dots,n\}$,
  all~$t\geq\tau$, and   all~$a\in H$.

\noindent Case 3.
Suppose that~$x_j(0)=0$ for some~$j$. Then there exists a \emph{minimal} index~$i$ such that~$x_i(0)=0$.
If~$i=n$ then~\eqref{eq:rfmeo_ode} yields
\begin{align*}
					 \dot x_{n}(0)&= \lambda_{n-1} x_{n-1}(0)-\lambda_{n} x_{n} (0)\\
					         &= \lambda_{n-1} x_{n-1} (0).
\end{align*}
By the definition of~$i$, $x_{n-1}(0)>0$ and thus~$x_{n}(0^+)>0$.

Now suppose that~$i=n-1$. Then~\eqref{eq:rfmeo_ode} yields
\begin{align*}
				\dot x_{n-1} (0)&= \lambda_{n-2} x_{n-2} (0)(1-y_{n+\ell-2}(0))-\lambda_{n-1} x_{n-1}(0)\\
					         &= \lambda_{n-2} x_{n-2}(0) (1-y_{n+\ell-2}(0)).
\end{align*}
By the definition of~$i$, $x_{n-2}(0)>0$.
If~$ \ell>2$ then~$1-y_{n+\ell-2}(0)=1$, and thus~$x_{n-1}(0^+)>0$.
If~$ \ell\leq 2$ then~$1-y_{n+\ell-2  }(0)=1-y_{n } (0)=1-x_n(0)-x_{n-1}(0)=1-x_n(0)$.
Thus, if~$x_n (0)<1$ then~$x_{n-1}(0^+)>0$. Consider the case~$x_n(0)=1$. Then~$y_n(0)=x_{n-1}(0)+x_n(0)=1$, so
\begin{align*}
					\dot y_{n}(0)&= \lambda_{n-\ell} x_{n-\ell}(0) (1-y_{n}(0))-\lambda_{n} x_{n}(0) \\
					         &= -\lambda_n     .
\end{align*}
This means that~$y_n(0^+)<1$, so again we conclude that~$x_{n-1}(0^+)>0$.

Continuing in this fashion shows that if~$x_j(0)=0$ for some~$j$ then~$x_j(0^+)>0$.
The analysis in all the other relevant cases is very similar, and thus omitted.~\IEEEQED

{\sl Proof of Prop.~\ref{prop:persist}.}
This follows from the fact that~$H$ is compact, convex and with a   repelling boundary; see~\cite[Thm.~2]{cast_book}
(see also~\cite{3gen_cont_automatica}).~\IEEEQED

{\sl Proof of Prop.~\ref{prop:weak_cont}.}
Pick~$\varepsilon, \tau>0$ and~$a,b\in H$. By Prop.~\ref{prop:persist},
 there exists~$\delta=\delta ( \tau  ) \in (0,1/2)$ such that for all~$i $ and all~$t\geq \tau$,
\be\label{eq:unbo}
			\delta \leq x_i(t),y_i(t) \leq 1-\delta.
\ee
Write the~$q_j$s in~\eqref{eq:rfmeo_ode} as
\begin{align*}
  q_j(x)&=\lambda_j x_j (1-y_{j+\ell}) \\
				&			=\eta_j x_j(1-x_{j+1}),
\end{align*}
where
$
			\eta_j (t):=\lambda_j \frac { 1-y_{j+\ell} (t)}{  1-x_{j+1}(t) } .
$
Note that~\eqref{eq:unbo} implies that
\be\label{eq:boiuns}
				0<\lambda_j \frac{\delta}{1-\delta} \leq \eta_j(t) \leq  \lambda_j \frac{1-\delta}{ \delta} <\infty
\ee
for all~$j$ and all~$t\geq \tau$.
Using this notation, the {\model} in~\eqref{eq:rfmeo_ode}
can be written as the time-varying system
\[
			\dot x_i = \eta_{j-1} x_{j-1}(1-x_{j})-\eta_j x_j(1-x_{j+1}).
\]
This means that for all~$t\geq \tau$
the {\model} can be interpreted   as an~RFM with time-varying transition rates~$\eta_j(t)$
that, by~\eqref{eq:boiuns},   are uniformly  bounded and
uniformly separated from zero
 for all~$t\geq \tau$.
Now the  results in~\cite{RFM_entrain} imply that
there exists~$\gamma:=\gamma(\varepsilon)$ such that after time~$\tau$
the solutions are
 contractive with overshoot~$(1+\varepsilon)$, and this completes the proof.~\IEEEQED

{\sl Proof of Prop.~\ref{prop:R_mono}.}
Consider two {\model}s, both with the same dimension~$n$ and particle size~$\ell$. The first with rates $\LMD$, admits a steady-state density $e$, and a steady-state production rate $R$, and the second with rates $\TLMD$, admits a steady-state density $\tilde e$ and a steady-state production rate $\tilde R$. Assume that there exists an index $j\in\{0,\dots,n\}$, such that $\tilde \lambda_i=\lambda_i$ for all $i\ne j$, and
\be\label{eq:tildejj}
\tilde \lambda_j > \lambda_j.
\ee
We need to show that $\tilde R>R$.
Seeking a contradiction, assume that
\be\label{eq:RtildeR}
\tilde R \le R.
\ee
We start with the case~$j=n$. Combining~\eqref{eq:RtildeR},~\eqref{eq:tildejj} and~\eqref{eq:ss_e} implies
 that $\tilde e_n < e_n$, and $\tilde e_{n-k}\le e_{n-k}$, $k=1,\dots,\ell-1$. This means that $\tilde y_n < y_n$, and combining this with~\eqref{eq:RtildeR} and~\eqref{eq:ss_e} implies that $\tilde e_{n-\ell} < e_{n-\ell}$, and so $\tilde y_{n-1} < y_{n-1}$. Continuing in this way yields $\tilde e_j < e_j$, $j=1,\dots,n-\ell$. In particular, $\tilde e_1+\dots + \tilde e_\ell <  e_1+\dots +  e_\ell$, and using~\eqref{eq:ss_e} results in $\tilde R > R$. This contradicts~\eqref{eq:RtildeR}, and so we conclude that $\tilde R > R$ in the case where $\tilde \lambda_n > \lambda_n$.

Using the same approach for any $j\in\{0,\dots,n\}$, while combining the assumption in~\eqref{eq:RtildeR} with~\eqref{eq:tildejj} and~\eqref{eq:ss_e}, yields
\begin{align}\label{eq:alltildey}
\tilde y_k &\le y_k,\quad k=j+\ell,\dots,n, \nonumber \\
\tilde y_k &< y_k,\quad k=\ell,\dots,j+\ell-1.
\end{align}
If $j>0$  then using $k=\ell$ in~\eqref{eq:alltildey} yields $\tilde e_1+\dots + \tilde e_\ell <  e_1+\dots +  e_\ell$, thus $\tilde R > R$, contradicting~\eqref{eq:RtildeR}. If~$j=0$   then using $k=\ell$ in~\eqref{eq:alltildey} yields $\tilde e_1+\dots + \tilde e_\ell \le e_1+\dots +  e_\ell$, but since $\tilde \lambda_0 > \lambda_0$, this again yields~$\tilde R > R$, contradicting~\eqref{eq:RtildeR}. We conclude that $\tilde R > R$.~\IEEEQED

{\sl Proof of Prop.~\ref{prop:thrfmeo_e}.}
Consider~\eqref{eq:ss_e} with $\lambda_0=\cdots=\lambda_n$. Then
\be\label{eq:th_en}
e_{n-\ell+1}=\cdots=e_n.
\ee
Since $e_{n-\ell}(1-e_{n-\ell+1}-\cdots -e_{n})=e_{n-\ell}(1-z_n)=e_{n-\ell+1}$, and $z_{n}\in(0,1)$, it follows that
\be\label{eq:th_enl}
e_{n-\ell}>e_{n-\ell+1},
\ee
and combining this with~\eqref{eq:th_en} implies that
\be\label{eq:th_zn}
z_{n-1}>z_n.
\ee
Now, since $e_{n-\ell-1}(1-e_{n-\ell}-\cdots -e_{n-1})=e_{n-\ell-1}(1-z_{n-1})=e_{n-\ell}(1-z_n)$, using~\eqref{eq:th_zn} and the fact that $z_{n-1},z_n\in(0,1)$ imply that $e_{n-\ell-1}>e_{n-\ell}$ and thus $z_{n-2}>z_{n-1}$. Continuing in this way completes the proof.~\IEEEQED

 {\sl Proof of Prop.~\ref{prop:R_dec_l}.}
Let $e$ $[ \bar e]$   denote the steady-state reader density in the {\model} [RFM].
 We need to show that $\bar R > R$.
Seeking a contradiction, assume that
\be\label{eq:propRdeca}
\bar R \le R.
\ee
Combining this with~\eqref{eq:ss_e} for both the {\model} with particle size $\ell$
and with particle size one (i.e. the RFM),   it follows that $\lambda_0(1-\bar e_1 )
\le \lambda_0( 1- e_1-\dots-e_\ell)$, thus
\[
\bar e_1 \geq  e_1+\dots+e_\ell,
\]
and since $e\in\Int(H)$ this yields
\be\label{eq:propRdec1}
\bar e_1 > e_1.
\ee
Using~\eqref{eq:ss_e},~\eqref{eq:propRdeca}, and~\eqref{eq:propRdec1}, it follows that
\[
\bar e_2 > e_2+\dots+e_{\ell+1},
\]
and since $e\in\Int(H)$ this yields
\[
\bar e_2 > e_2.
\]
Continuing in this way yields
\be\label{eq:propRdecj}
\bar e_j > e_j +\dots+e_{j+\ell-1},\quad j=2,\dots,n-\ell+1,
\ee
so in particular,
\be\label{eq:whaytt}
\bar e_{n-\ell+1}  > e_{n-\ell+1}  +\dots+e_{n} .
\ee
On the other-hand using~\eqref{eq:propRdeca} and comparing the last $\ell$ equations in~\eqref{eq:ss_e}
for both the {\model} with particle size $\ell$
and with particle size one (i.e. the RFM), yields
\begin{align}\label{eq:propRdecl}
\bar e_{n-\ell+1}(1-\bar e_{n-\ell+2}) &\le e_{n-\ell+1}, \nonumber \\
\bar e_{n-\ell+2}(1-\bar e_{n-\ell+3}) &\le e_{n-\ell+2}, \nonumber \\
\dots \nonumber  \\
\bar e_{n-1}(1-\bar e_n) &\le e_{n-1}, \nonumber \\
\bar e_n &\le e_n.
\end{align}
Now combining~\eqref{eq:propRdecl} with~\eqref{eq:whaytt} yields
\be\label{eq:propRdec0}
  \bar e_{n-\ell+2}(1-\bar e_{n-\ell+1}) + \bar e_{n-\ell+3}(1-\bar e_{n-\ell+2})+\dots+\bar e_n(1-\bar e_{n-1})<0.
\ee
However, since $\bar e \in \Int(H)$, the term on the left-hand side here must be strictly positive.
This contradiction completes the proof.~\IEEEQED


\section*{\updt{Appendix~B:  {\model} as a Mean-Field Approximation of TASEPEO}}
\updt{In this appendix, we show how the {\model} can be derived from~TASEPEO. We use a 
 notation  that is standard    in the~TASEPEO literature.

Consider TASEPEO with $N$ sites, rates $\mu$ defined in~\eqref{eq:tasep_rates},
 extended object size $\ell$, and under the assumption
that  the reader is located at the left-most site of the object. Following   MacDonald et. al.~\cite{MacDonald1968} (see also~\cite{macdonald1969concerning}) the current from site~$i$ to site~$i+1$ at time~$t$ is given by (for simplicity we ignore boundary cases):\footnote{\updt{Note that in~\cite{MacDonald1968} the reader is defined to be in the~\emph{right-most site} of the object, and thus there the current is proportional to the probability that site $i$ has a reader and site $i+1$ is empty.}}
\begin{align}\label{eq:tasep_J}
J_{i\to i+1}(t)&=\gamma_i \pr(\text{site }  i \text{ has a reader  and site } i+\ell \text{ is empty}) \nonumber \\
&=\gamma_i \pr(\text{site }  i \text{ has a reader}) \pr(\text{site } i+\ell \text{ is empty } |  \text{ site }  i \text{ has a reader}),
\end{align}
where $\pr(a)$ [$\pr( a | b )$] denotes the probability of  event $a$ [the conditional probability of   event $a$ given   event $b$] at time $t$. 
Since the conditional probability in~\eqref{eq:tasep_J} is difficult to estimate, we apply what~\cite{PhysRevE.76.051113}
calls a naive mean-field approximation,  and replace~\eqref{eq:tasep_J} by:
\begin{align}\label{eq:tasep_L_mf}
J_{i\to i+1}(t)&=\gamma_i \pr(\text{site }  i \text{ has a reader}) \pr(\text{site } i+\ell \text{ is empty}) \nonumber \\
&=\gamma_i\pr(\text{site }  i \text{ has a reader}) \left(1-\sum_{k=0}^{\ell-1} \pr(\text{site }  i+\ell-k \text{ has a reader})\right).
\end{align}
We approximate the probabilities above by averaging the binary reader occupancies over an ensemble of~TASEPEO systems, i.e. we replace $\pr(\text{site }  i \text{ has a reader}) $ by $\rho^r_i(t):=\langle  r_i(t) \rangle$, where $r_i(t)\in\{0,1\}$ is the reader occupancy at site $i$ at time $t$, and the operator $\langle \rangle$ denotes an average over the ensemble. This yields
\be\label{eq:tasep_Jii1}
J_{i\to i+1}(t) = \gamma_i \rho^r_i(t) \left(1-\sum_{k=0}^{\ell-1} \rho^r_{i+\ell-k}(t)\right ).
\ee
The change in the average reader occupancy at site $i$ at time $t$ is given by~\cite{MacDonald1968}:
\be\label{eq:tasep_drho}
\frac{d}{dt} \rho^r_i(t)=J_{i-1\to i}(t)-J_{i\to i+1}(t).
\ee
Introducing  the notation
$
x_i(t):=\rho^r_i(t)$ and 
$\lambda_i:= \gamma_i ,
$
we see that $J_{i\to i+1}(t)$ corresponds to $q_i(x)$ in~\eqref{eq:rfmeo_flow}, and~\eqref{eq:tasep_drho} 
corresponds to~\eqref{eq:rfmeo_ode} (see~\eqref{eq:y}). Thus, we obtained the~{\model}. 
 In particular, the case $\ell=1$ in~\eqref{eq:tasep_Jii1} corresponds to the  dynamical equations of the~RFM
(see~\eqref{eq:rfm_all}).

At steady-state, we expect every~$\rho^r_i(t)$ in~TASEPEO to converge to, say, $\rho^r_i$, and then
the currents between any two consecutive sites are all equal (but we  
 are not aware of any rigorous proof of   convergence in~TASEPEO). 
The derivation above (including  the boundary cases as well~\cite{macdonald1969concerning,PhysRevE.76.051113})
shows that  the steady-state current satisfies:
\begin{align}\label{eq:J_tasep_mf}
J&=\alpha(1-\sum_{k=0}^{\ell-1} \rho^r_{\ell-k}) \nonumber \\
&=\gamma_i \rho^r_i(1-\sum_{k=0}^{\ell-1} \rho^r_{i+\ell-k}) &\text{ for all } & 1\le i \le N-\ell \nonumber \\
&=\gamma_i \rho^r_i &\text{ for all }  & N-\ell+1 \le i \le N-1 \nonumber \\
&=\beta \rho^r_N.
\end{align}
If we use the notation~$e_i:=\rho^r_i$, $\lambda_0:=\alpha$, and~$\lambda_n:=\beta$  then this is 
just the steady-state equation  of~{\model} given in~\eqref{eq:ss_e}. 
}
\bibliographystyle{IEEEtranS}  

\end{document}